\documentclass[usenatbib,usegraphicx,usedcolumn]{mn2e}
\usepackage{times,amssymb,amsmath}

\title[TFR as function of environment]{The Tully-Fisher Relation for 25,000 SDSS Galaxies as Function of Environment}
\author[P. Mocz et al.]{
P. Mocz$^{1,2}$\thanks{E-mail: pmocz@fas.harvard.edu (PM); agreen@astro.swin.edu.au (AG); maximus.malacari@adelaide.edu.au (MM); karl@astro.swin.edu.au (KG)}, 
A. Green$^{1}$\footnotemark[1],
M. Malacari$^{1,3}$\footnotemark[1],
K. Glazebrook$^{1}$\footnotemark[1]\\
$^{1}$Centre for Astrophysics and Supercomputing, Swinburne University, PO Box 218, Hawthorn, VIC 3122, Australia\\
$^{2}$Department of Astronomy, Harvard University, 60 Garden Street, Cambridge, MA 02138, USA\\
$^{3}$Department of Physics, University of Adelaide, Adelaide, SA 5005, Australia}

\begin{document}

\date{subm. to MNRAS, 30 August 2011, Accepted}

\pagerange{\pageref{firstpage}--\pageref{lastpage}} \pubyear{2011}

\maketitle

\label{firstpage}

\begin{abstract}
  We construct Tully-Fisher relationships (TFRs) in the $u$, $g$, $r$,
  $i$ and $z$ bands and stellar mass TFRs (smTFRs) for a sample of
  $25,698$ late spiral type galaxies (with $0.045<z<0.085$) from the
  Sloan Digital Sky Survey (SDSS) and study the effects of environment
  on the relation. We use SDSS-measured Balmer emission line widths,
  $v_{\rm FWHM}$, as a proxy for disc circular velocity, $v_{\rm
    circ}$. A priori it is not clear whether we can construct accurate
  TFRs given the small $3''$ diameter of the fibres used for SDSS
  spectroscopic measurements. However, we show by modelling the
  H$\alpha$ emission profile as observed through a $3''$ aperture that
  for galaxies at appropriate redshifts ($z>0.045$) the fibres sample
  enough of the disc to obtain a linear relationship between $v_{\rm
    FWHM}$ and $v_{\rm circ}$, allowing us to obtain a TFR and to
  investigate dependence on other variables. We also develop a
  methodology for distinguishing between astrophysical and sample bias
  in the fibre TFR trends. We observe the well-known steepening of the
  TFR in redder bands in our sample. We divide the sample of galaxies
  into four equal groups using projected neighbour density ($\Sigma$)
  quartiles and find no significant dependence on environment,
  extending previous work to a wider range of environments and a much 
  larger sample. Having demonstrated that we can construct SDSS-based 
  TFRs is very useful for future applications because of the large 
  sample size available.
\end{abstract}


\section{Introduction}\label{sec:intro}

The TFR is an observed correlation between the luminosity and disc
circular velocity, $v_{\rm circ}$, of disc type galaxies
\citep{1977A&A....54..661T}. This fundamental empirical relationship
reflects important physics in galaxy formation
(e.g. \citealt{1998MNRAS.295..319M}) and serves as a distance
indicator to galaxies (e.g. \citealt{1977A&A....54..661T},
\citealt{2000ApJ...533..744T} and \citealt{2001ApJ...553...47F}). The
luminosity of a galaxy is proportional to the stellar mass, while the
rotation curve (RC) circular velocity is determined primarily by the
dark matter halo (which is significantly more massive), so the TFR is
essentially a relation between the dark matter component and the
luminous baryonic component of a galaxy.

The TFR was discovered by \cite{1977A&A....54..661T} who measured
luminosity versus $21$~cm neutral hydrogen emission line width with
the $91$~m National Radio Astronomy Observatory (NRAO) for a sample of
$10$ spiral galaxies. Using $21$~cm data is advantageous because of
the ability of a radio telescope to sample the neutral hydrogen to
very large galactic radii, obtaining data up to the plateau of the
RC. Since its discovery, the TFR has been investigated using other
measures of disc circular velocity as well, such as optical H$\alpha$
emission with long-slit spectroscopy
(e.g. \citealt{1997AJ....114.2402C}, \citealt{2002AJ....123.2358K} and
\citealt{2007AJ....134..945P}).  TFRs have also been reconstructed
from optical one-dimensional spectra (integrated line-of-sight
velocity widths) for galaxies up to $z\sim 1$
\citep{2006ApJ...653.1027W, 2006ApJ...653.1049W}.

There have been many studies of the TFR, each with its own
  selection criteria, but generally falling in two broad camps.%
  The first camp is those interested in the physical processes which
  give rise to the TFR, e.g. \cite{2007AJ....134..945P,
    2002AJ....123.2358K, 2001ApJ...550..212B, 2010A&A...510A..68P, 1997AJ....114.2402C} and
  many others. The introduction to
  \citealt{1997AJ....114.2402C} gives an excellent review. These
  surveys tend to have broader selections, for example,
  \cite{1997AJ....114.2402C} select bright ($m_B < 15.5$), large
  (diameter $\geq 4\arcmin$), inclined ($55\degr < i < 75\degr$)
  galaxies from the UGC, eliminating only galaxies with large dust
  extinction or peculiar/interacting morphologies
  \citep{1993ApJ...412L..51C}. Scatter in the TFR remains a
  large uncertainty in its use as a distance indicator, so physical
  explanations of this scatter such as that of
  \cite{2002AJ....123.2358K} are a key area of research. %
  The second camp is interested in the TFR as a empirical law for
  measuring distances, e.g. \cite{1977A&A....54..661T,
    1992ApJS...81..413M, 1996ApJS..107...97M, 2007ApJS..172..599S} and
  many others. The use of the TFR as a distance indicator is reviewed
  in \cite{1995PhR...261..271S} and \cite{2000ApJ...533..744T}. To
  minimise scatter in distance measurements, these studies tend to
  have narrower selection criteria, preferentially selecting late-type
  spirals (Sb -- Sd) and luminosity pass-bands where intrinsic scatter
  is minimised. Despite differences, these two camps are closely
  linked: a better physical understanding of the TFR provides a better
  calibration of disk galaxy luminosities and hence more accurate
  distances.

  We will make extensive use of one such study here for
  calibration and comparison: the paper by
  \cite{2007AJ....134..945P}. That work investigates a broadly
  selected input sample of $234$ galaxies drawn from the SDSS with
  $-22 < M_r < -18.5$ to study the intrinsic scatter in the TFR.
They obtain reliable H$\alpha$ RCs from long-slit spectroscopic
measurements with the Calar Alto $3.5$~m telescope and the MDM $2.4$~m
telescope for $162$ galaxies from the sample. This broadly selected
sample does, however, contain typical magnitudes of $M_r < -18.5$ and
$r$-band scale heights of $8''$, meaning it only samples the brightest
and largest low-$z$ galaxies in the SDSS. They find the slope of the
TFR systematically steepens from $-5.5\pm0.2$~mag~$(\log_{10} {\rm
  km}~{\rm s}^{-1})^{-1}$ in the $g$ band to
$-6.6\pm0.2$~mag~$(\log_{10} {\rm km}~{\rm s}^{-1})^{-1}$ in the $z$
band. The intrinsic scatter of the TFR is $0.4$~mag in the $g$, $r$,
$i$ and $z$ bands, and is attributed to being driven largely by the
variations in the ratio of the dark to luminous matter within the disc
galaxy population because correlations of the TFR with galaxy
properties such as colour, size and morphology were found to be
weak. The slope of the TFR derived by
  \citeauthor{2007AJ....134..945P} is shallower than contemporary
  studies, which they describe as an effect of Malmquist-type biases
  for which they include no correction. We have chosen this sample for
  comparison here because it is based entirely on SDSS data.

The large intrinsic scatter in the TFR suggests that the TFR may
depend on properties external to the galaxy, such as the environment
(i.e., the local number density of galaxies). Environment is
  known to have a strong impact on galaxy evolution, and, in theory,
  could have an effect on the TFR. For example, if galaxies in a
higher density environment undergo accelerated star formation due to
tidal interactions or mergers, then they may be overly luminous for
their rotational velocity and the TFR will be altered. Or, a galaxy
falling into a cluster can have a significant amount of its gas
stripped \citep{1999MNRAS.308..947A}, which can alter the
TFR. Studying the TFR as a function of environment hence has
implications for galaxy evolution. \cite{2006ApJ...653..861M}
conducted two simple tests for an effect of environment on the TFR and 
found no variation across a range of cluster environments in a sample 
of 5000 galaxies as a function of clusters.

The goal of this paper is the measure the TFR for a large sample
($\sim25,000$) of Sloan Digital Sky Survey \citep[SDSS;][]{2000AJ....120.1579Y} galaxies as a function of environment with very
little sample selection criteria except for a colour cut, which will
select mostly late disc type galaxies from SDSS. The TFR has not been
investigated as a function of environment statistically before,
although comparisons have been made between cluster and field TFRs as
well as the TFRs for different clusters (see
\citealt{1992ApJ...387...47P}, \citealt{2003AJ....126.2622H},
\citealt{2006ApJ...653..861M}, \citealt{2007ApJS..172..599S},
\citealt{2009ApJS..182..474S} and references
therein). \cite{2003AJ....126.2622H} gather data on $13$ Coma Cluster
S0 galaxies and $8$ Virgo Cluster S0 galaxies to compare their TFRs
with the TFRs for late-type spirals and field S0s. They find no
difference between the TFRs for field and cluster S0s, and only a
small offset from the TFR for late-type spirals. However, the sample
size is small. \cite{2006ApJ...653..861M} studies $\sim5000$ field and
cluster spiral galaxies that have either $21$~cm or H$\alpha$ line
width measurements to investigate the $i$ band TFR. The sample comes
from several datasets from the 1990s as well as significant amounts of
new data, which are carefully homogenised, and is presented
in \citet{2007ApJS..172..599S} (see also the erratum
\citealt{2009ApJS..182..474S}). \cite{2006ApJ...653..861M} find a
morphological dependence on the TFR: earlier type spirals are observed
to be on average dimmer than later type spirals at a fixed velocity
width, which results in a shallowing of the TFR slope for earlier type
spirals. This is attributed by \cite{2006ApJ...653..861M} to be the
reason why field versus cluster TFRs may be different (since
early-type spirals are more likely to be found in over-dense
regions). However, \cite{2006ApJ...653..861M} find no difference in
the slopes, zero points, and scatters of the TFRs for $31$ nearby
clusters and groups, and see no trend in the TFR residuals with a
galaxy's projected distance from their cluster centre when
investigating the impact of environment.

In the present work, we directly investigate the TFR as a function of the local density $\Sigma$, determined from the projected distances to the fourth and fifth nearest neighbours, ranging from voids to values typical for clusters for a broad sample of SDSS galaxies. The projected densities have been calculated as in \cite{2006MNRAS.373..469B}, which have been applied in previous studies to examine the environmental dependence of a number of galaxy properties for large samples of SDSS galaxies. \cite{2006MNRAS.373..469B} use $\Sigma$ to find that the fraction of red galaxies increases with increasing $\Sigma$ as well as galaxy mass. \cite{2007MNRAS.382..801M} investigate the stellar mass--metallicity (gas-phase oxygen abundance) relation in star-forming galaxies as a function of projected neighbour density and find no dependence of environment on the relation, and conclude that the evolution of the galaxies is largely independent of their environments. However, \cite{2007MNRAS.382..801M} do find marginal increase in the chemical enrichment level at a fixed stellar mass in denser environments, suggesting that environments can play a moderate role in galaxy evolution.

The organization of the paper is as follows. In Section~\ref{sec:why}
we show by modelling aperture-biased H$\alpha$ line profiles why we
can expect to obtain a TFR using SDSS-observed line widths. In
Section~\ref{sec:data} we discuss the data and sample selection used
to construct our TFRs. In Section~\ref{sec:results} we investigate the
environmental effects on the TFRs in our sample (and find no strong
dependence on environment). We find our TFRs are consistent with the
TFRs of \cite{2007AJ....134..945P}. In Section~\ref{sec:disc} we
discuss the implications of our results and our conclusions.  The
model details are found in the Appendix, as well as discussion of
effects which may affect the sensitivity of our method (seeing,
Malmquist bias, rotation curve shape).  We use standard cosmological
parameters of $H_0=71$~km~s$^{-1}$~Mpc$^{-1}$, $\Omega_{\rm m} = 0.27$
and $\Omega_\Lambda = 0.73$ throughout this paper. All errors are
quoted at the $95$ per cent confidence level unless otherwise stated.

\section{Why we can expect to obtain a TFR with SDSS line widths}\label{sec:why}

We investigate a large sample of galaxies from the SDSS
\citep{2000AJ....120.1579Y}. We use as a measure of the circular disc
velocity the SDSS-observed, inclination-corrected Balmer emission
full-width half-maximum (FWHM) velocity width, $v_{\rm FWHM}$ (which
have been reprocessed in the Max Planck for Astronomy and Johns
Hopkins University (MPA-JHU) spectroscopic re-analysis).  These widths
suffer from aperture bias since SDSS galaxy spectra are obtained
through a $3''$ diameter fibre, which means that the fibres do not
cover enough of the galaxy all the way out to the maximal disc
circular velocity $v_{\rm circ}$ to provide direct measurements of
$v_{\rm circ}$. However, we may still be able to obtain a TFR if
$v_{\rm circ}$ is correlated with the observed, aperture-biased
$v_{\rm FWHM}$. A number of factors may aid in the existence of a
relationship between $v_{\rm circ}$ and $v_{\rm FWHM}$. More luminous
galaxies are farther away and hence smaller in angular size so the
fibres cover more of the galaxy. Less luminous galaxies are nearby but
physically smaller (in addition, by the TFR, the value for $v_{\rm
  circ}$ is also smaller which will also decrease the discrepancy
between the actual $v_{\rm circ}$ and its value implied by $v_{\rm
  FWHM}$).  Seeing (typical FWHM is $\sim1.4''$) may potentially help
by increasing the observed FWHM, bringing it closer to the actual FWHM
as would be observed by a large enough fibre, since some light
entering the fibre from the galaxy comes from beyond the $3''$
aperture projected onto the galaxy (although more light, from within
the $3''$ aperture projection, is scattered out).

\begin{figure}
\centering
\includegraphics[width=0.47\textwidth]{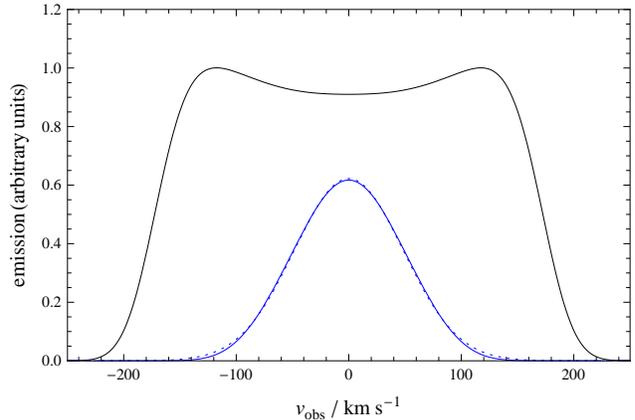}
\caption{Example H$\alpha$ emission profile calculated from our disc model for parameters $i=80^\circ$, $h=3''$, $r_{\rm t}=2$ and $v_{\rm circ}=200$~km~s$^{-1}$ (model details are found in Appendix~\ref{sec:model}). The black solid line shows the full profile without aperture effects. The blue solid line shows how the same profile would appear if observed with a $3''$ aperture and $1.4''$ seeing, like in SDSS observations. A Gaussian is fit to the aperture-biased profile, shown as a blue dashed line, which nearly coincides with the actual aperture-biased profile. The full profile, on the other hand, may show a double-peaked characteristic (depending on model parameters) as in this case.}
\label{fig:TP}
\end{figure}

Here, we develop a model for the observed H$\alpha$ profile of a disc galaxy as measured through a $3''$ aperture, accounting for $1.4''$ FWHM seeing. The model assumes an infinitely thin inclined disc with an exponential surface brightness profile, a radial velocity profile characterised by an asymptotic circular velocity and a turn-over radius, and constant velocity dispersion across the galaxy. We then use our model to create a synthetic population with model parameters chosen according to distributions inferred from the SDSS and the local TFR. We investigate the relation between the true $v_{\rm circ}$ and the observed inclination-corrected $v_{\rm FWHM}$ as a function of redshift to learn at which redshifts it is possible to construct reliable TFRs using SDSS line widths. The details of the model are found in Appendix~\ref{sec:model}. An example H$\alpha$ emission profile with and without the SDSS $3''$ aperture bias is presented in Figure~\ref{fig:TP}. The modelling performed here in Section~\ref{sec:why} is used to inform our final sample selection (Section~\ref{sec:data}) of disc galaxies from SDSS Data Release Eight (DR 8) \citep{2011ApJS..193...29A} for studying the TFR.

Previous studies in the literature have performed similar simulations of line widths as observed by fibre spectra to understand their observed distribution. For example, \cite{1997MNRAS.285..779R} simulated the expected distribution of (observable) [O{\sc II}] emission line widths in galaxies (for a much smaller sample than ours) as would be seen by the AUTOFIB fibre spectrograph on the Anglo-Australian Telescope (AAT). They assumed various galaxy circular rotation speeds, and included effects of random viewing angles, clumpy line emission, finite fibre aperture and internal dust extinction on the profile of the emission line.

\subsection{Relation between $v_{\rm circ}$ and $v_{\rm FWHM}$}\label{sec:why2}

The free parameters in our model for H$\alpha$ emission are
inclination $i$, exponential scale radius $h$, dynamic turn-over
radius $r_{\rm t}$ and asymptotic disc velocity $v_{\rm
  circ}$. Assuming distributions for these parameters that are typical
for SDSS disc galaxies allows us to test for a correlation between
$v_{\rm circ}$ and $v_{\rm FWHM}$. A reasonable correlation would
allow for the construction of a TFR using SDSS-measured line
widths. The values for $v_{\rm FWHM}$ we obtain are from the FWHM of
the model profiles and inclination corrected (i.e., divided by
$\sin(i)$). Inclination-correction reduces the scatter between $v_{\rm
  circ}$ and $v_{\rm FWHM}$.

We choose inclinations $i$ between $55^{\circ}$ and
$90^{\circ}$. Assuming random orientation, this is achieved by drawing
a value for $\sin(i)$ from a uniform distribution between
$\sin(55^{\circ})$ and $\sin(90^{\circ})$. The reason for this choice
of inclinations is that in our sample selection, we will want to only
analyse galaxies that are more than $55^\circ$ from face on, otherwise the
projected line-of-sight velocities are too small.

\begin{figure*}
\centering
\includegraphics[width=\textwidth]{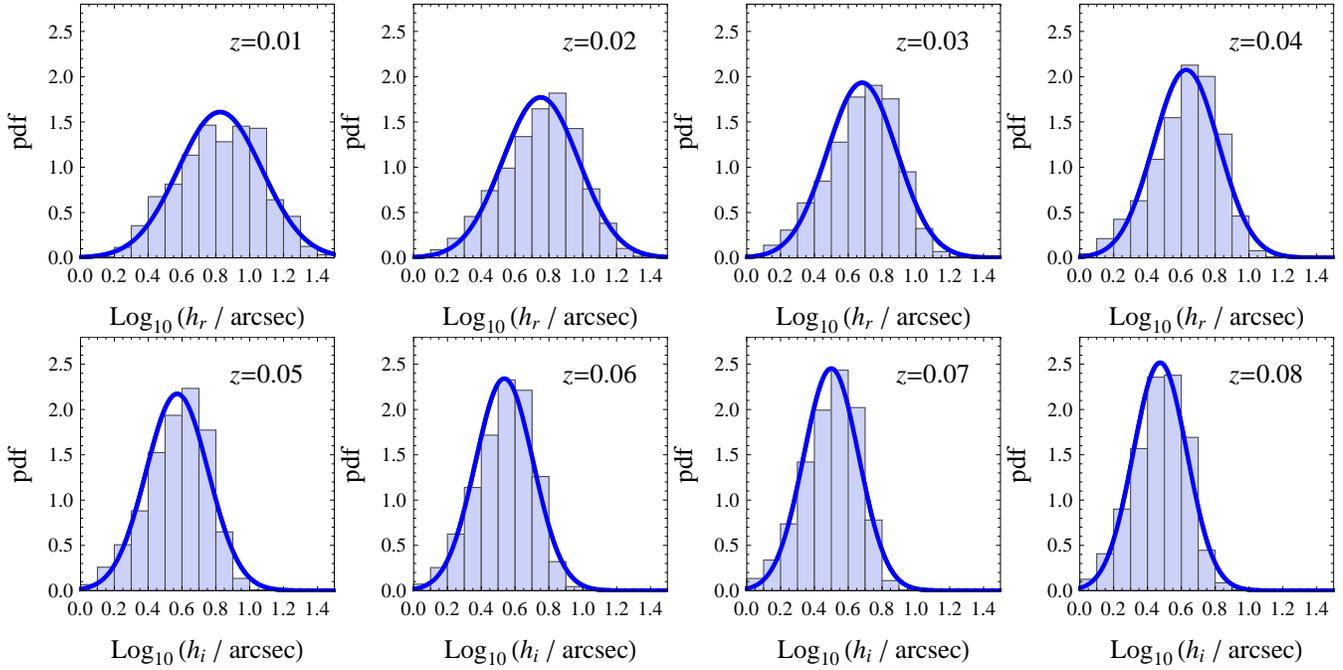}
\caption{Histograms showing the SDSS probability distributions of exponential scale lengths, fitted with Gaussians, used to create our synthetic population.}
\label{fig:exprhist}
\end{figure*}

Establishing distributions for $h$, $r_{\rm t}$ and $v_{\rm circ}$ is
more complicated. The distributions are a strong function of redshift
because farther galaxies appear smaller, and more luminous galaxies
can be seen farther away (and hence have higher $v_{\rm circ}$). We
establish the distribution from a sample of $44,276$ SDSS DR 8
galaxies, Sample A. The details of the selection of this Sample A are
described in Section~\ref{sec:data}. Sample A consists of galaxies
with redshifts $0.005<z<0.085$ and magnitudes $-22.5<M_r<-15$. We
divide the sample into groups with redshifts centered around
$z=0.01,0.02,...,0.08$ and interval widths $\Delta z = 0.01$. To
obtain a distribution for $h$ for a given redshift group, we assume
that the H$\alpha$ emission profile traces the profile of the emission
of the band in which the H$\alpha$ line is located. In the rest frame,
H$\alpha$ is at $656.5$~nm, which is in the $r$-band. Beyond $z\simeq
0.05$, the emission is redshifted to the $i$-band. We find that the
exponential scale lengths have approximately a log-normal
distribution, as shown in Figure~\ref{fig:exprhist}. The probability
distributions in each redshift bin are fitted with log-normal
distributions, with parameters listed in Table~\ref{tbl:synps}. These
distributions were used to construct our synthetic population.

\begin{table}
\centering
\caption{Gaussian probability distributions for synthetic population parameters}\label{tbl:synps}
\begin{tabular}{@{}ccccc@{}}
\hline\hline
\noalign{\smallskip}
 & \multicolumn{2}{c}{$\log_{10}(h/\rm{arcsec})$} & \multicolumn{2}{c}{$\log_{10} (v_{\rm circ} / \rm{km}~\rm{s}^{-1})$} \\
$z$-interval & $\mu$ & $\sigma$ & $\mu$ & $\sigma$ \\[2pt]
\hline 
\noalign{\smallskip}
$[0.005,0.015]$ & $0.82$ & $0.25$ & $1.63$ & $0.20$ \\
$[0.015,0.025]$ & $0.75$ & $0.23$ & $1.78$ & $0.16$ \\
$[0.025,0.035]$ & $0.68$ & $0.21$ & $1.88$ & $0.14$ \\
$[0.035,0.045]$ & $0.63$ & $0.19$ & $1.96$ & $0.12$ \\
$[0.045,0.055]$ & $0.57$ & $0.18$ & $2.02$ & $0.10$ \\
$[0.055,0.065]$ & $0.54$ & $0.17$ & $2.08$ & $0.09$ \\
$[0.065,0.075]$ & $0.50$ & $0.16$ & $2.12$ & $0.08$ \\
$[0.075,0.085]$ & $0.47$ & $0.16$ & $2.16$ & $0.07$ \\
\hline
\end{tabular}
\end{table}

\begin{figure}
\centering
\includegraphics[width=0.47\textwidth]{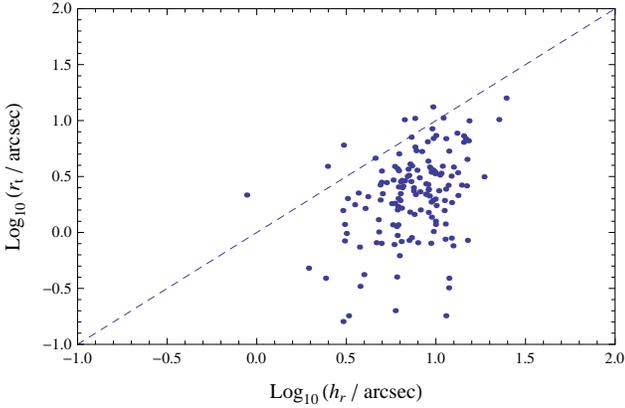}
\caption{Plot of $r_{\rm t}$ versus $h_r$ for a sample of SDSS selected galaxies from \protect\cite{2007AJ....134..945P}.}
\label{fig:rthr}
\end{figure}

Next, we wish to describe a distribution for the velocity turn-over
radius $r_{\rm t}$. This information is not measured by the SDSS. We
investigate the sample of $162$ SDSS selected local galaxies from
\cite{2007AJ....134..945P} with reliable H$\alpha$ RCs obtained from
long-slit spectroscopic measurements with the Calar Alto $3.5$~m
telescope and the MDM $2.4$~m telescope. We find no significant
correlation between $r_{\rm t}$ and SDSS variables such as magnitude
or $h_r$ (exponential scale radius in the $r$-band) which would help
us predict a value of $r_{\rm t}$. A plot of $r_{\rm t}$ versus $h_r$
is presented in Figure~\ref{fig:rthr}. In the
\cite{2007AJ....134..945P} sample, the mean value of $r_{\rm t}$ is
$0.5$~dex smaller than $h_r$ and has an approximately log-normal
distribution. Therefore, given a distribution of $h_r$, we assume a
log-normal distributions for $r_{\rm t}$ with mean $0.5$~dex smaller
than the $h_r$ mean and the same standard deviation.

\begin{figure*}
\centering
\includegraphics[width=\textwidth]{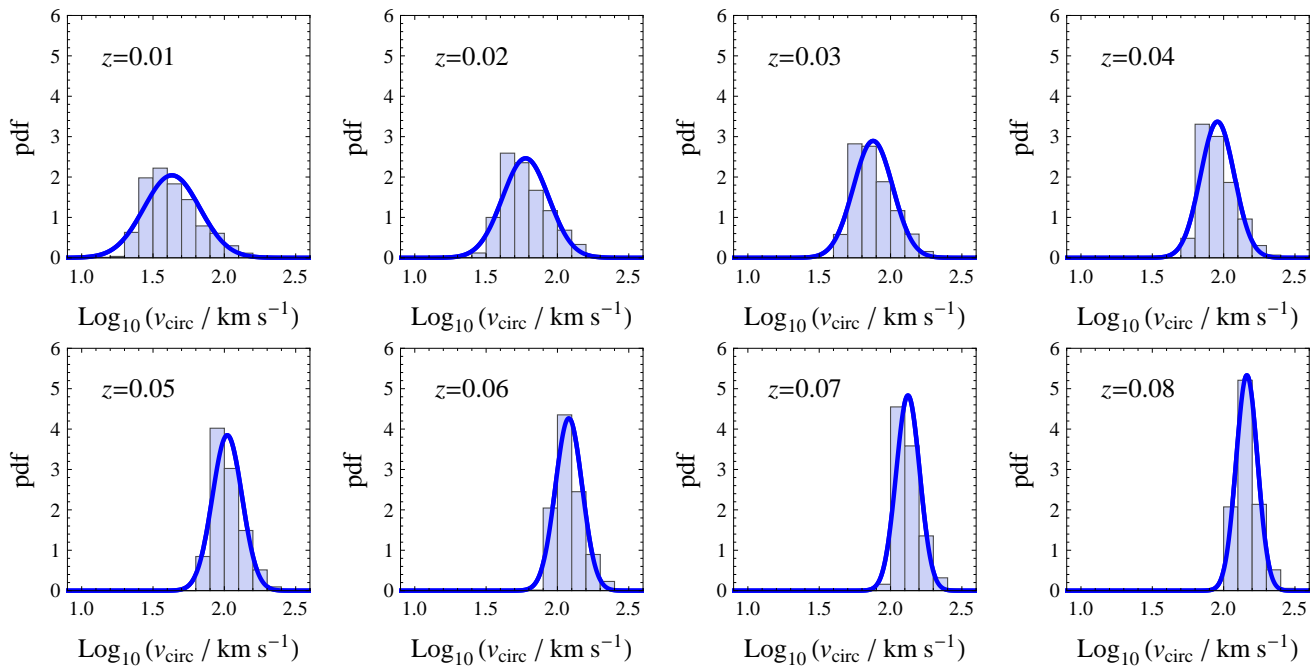}
\caption{Histograms showing the SDSS probability distributions of $v_{\rm circ}$ (derived assuming the local TFR from \citealt{2007AJ....134..945P}), fitted with Gaussians, used to create our synthetic population.}
\label{fig:Vcirc}
\end{figure*}

Finally, to obtain distributions for $v_{\rm circ}$, we use the
distributions of absolute $i$-band magnitudes $M_i$ and convert them
to $v_{\rm circ}$ using the local Tully Fisher relation found in
\cite{2007AJ....134..945P}:
\begin{eqnarray}
M_i &=& -6.321 (\log_{10}v_{80}-2.220) -21.390 \\
    &=& -6.321 \log_{10}v_{80}- 7.357 
\end{eqnarray}
where $v_{80}$ is the velocity (in km~s$^{-1}$) at a radius containing
$80$ per cent of the $i$-band flux, with $v_{80}\simeq v_{\rm circ}$
in the \cite{2007AJ....134..945P} sample. 
The distributions of $v_{\rm
  circ}$ for the different redshift bins are plotted in
Figure~\ref{fig:Vcirc}, which again are approximately log-normal, with
best-fit parameters listed in Table~\ref{tbl:synps}.

\begin{figure*}
\centering
\includegraphics[width=\textwidth]{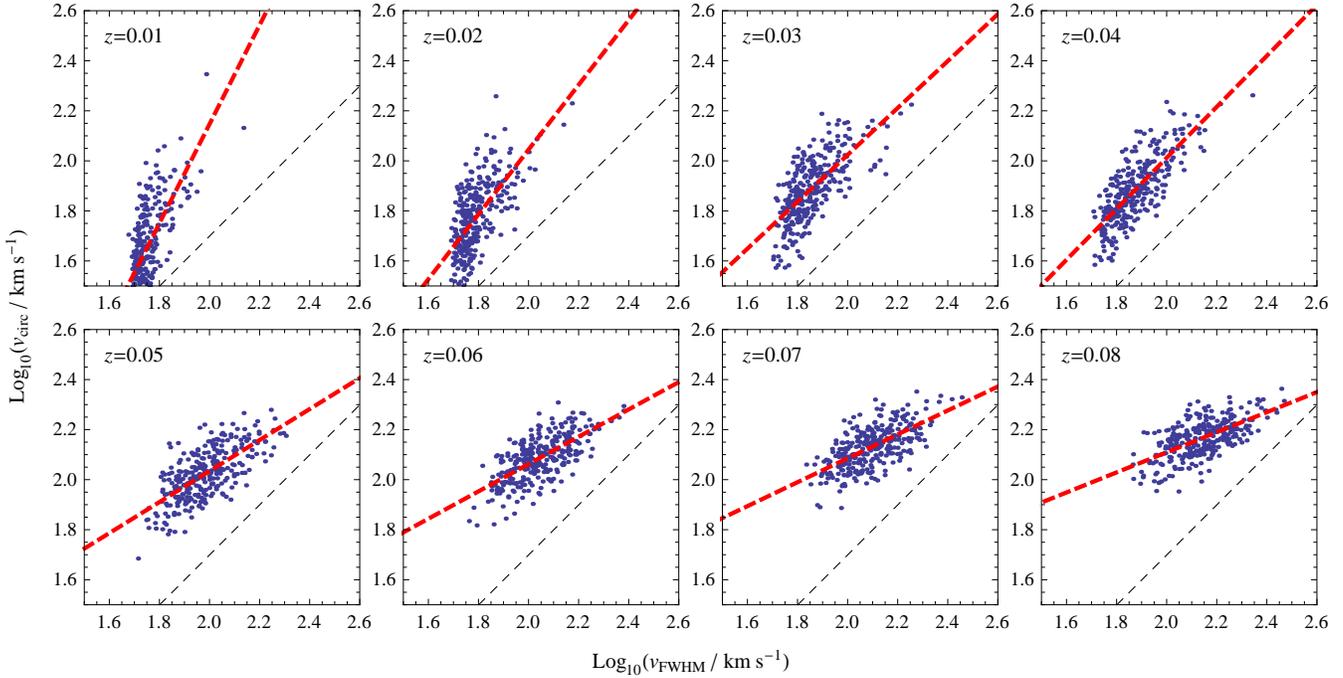}
\caption{The expected relation between $v_{\rm circ}$ and $v_{\rm FWHM}$ assuming our simple disc model, free parameters drawn according to SDSS distributions, and the local TFR. The $v_{\rm FWHM}$ are inclination-corrected. Since by the TFR $v_{\rm circ}$ corresponds to some magnitude $M$ (so the $y$-axes can be re-scaled as magnitudes), this plot shows how the shape of a scatter plot of absolute magnitude versus SDSS-measured velocity widths ($v_{\rm FWHM}$) looks like. Strong aperture bias is seen for galaxies at $z\leq 0.045$. The thick, red, dashed lines show the best-fit linear models to the correlation. The thin, black, dashed lines indicate below which no $v_{\rm FWHM}$'s should fall, given $v_{\rm circ}$ and assuming an arctan velocity profile.}
\label{fig:whyTF}
\end{figure*}

Using the above distributions for the free parameters, we construct a synthetic population of $300$ galaxies for each of the redshift bins using our disc model. We plot the relation between the actual $v_{\rm circ}$ and SDSS-measured, inclination corrected $v_{\rm FWHM}$ in Figure~\ref{fig:whyTF}. We see that at low redshifts ($z\leq 0.045$), where galaxies appear larger in angular size and hence we are probing only the central parts of the emission with the fibres, there is clear aperture-bias effects with the distribution of $v_{\rm circ}$ versus $v_{\rm FWHM}$ being steep and fan-shaped. At higher redshifts ($z\geq 0.045$), the relation can be modelled with a linear fit. 
At these redshifts, the residuals show no biasing pattern and the 
variance (sum of the squares of the distance to the fit divided by 
the number of points) has dropped significantly: the variance from 
the low through the high redshift bins are: 
$0.0147$, $0.0126$, $0.0086$, $0.0073$, $0.0055$, $0.0045$, $0.0041$, $0.0035$ 
respectively.
The scatter is still large in the relation due to aperture effects, however this does not prevent us from obtaining a good fit to the slope of the TFR (although this does make it difficult to use SDSS-measured line widths to study the intrinsic scatter in the TFR). 

The plots in Figure~\ref{fig:whyTF} can be thought of as indicating the expected shape of the SDSS line width-based TFR (i.e., functions of $v_{\rm FWHM}$), since an assumed TFR (as function of $v_{\rm circ}$) can be used to convert the $y$-axis to absolute magnitudes. The shape of the distribution is not expected to be linear if we select galaxies at all redshifts between $0.005$ and $0.085$, due to the aperture effects that become more prominent for lower $z$ galaxies (which appear larger in angular size). The low $z$ (and hence also dimmer) galaxies are expected to distinctly turn-off from the linear trend the higher $z$ galaxies demonstrate due to the aperture bias. Therefore, in order to construct and analyse the TFR for a sample of SDSS galaxies, we will choose only galaxies with $z\geq 0.045$.

  We have chosen the \cite{2007AJ....134..945P} relation because
  it is directly based on SDSS data, and therefore made for the
  simplest comparison. However, there are several caveats to this
  decision. The \cite{2007AJ....134..945P} relation is shallower than
  most current TFRs, primarily because Malmquist-type biases are not
  adequately addressed (as described in their Section~5.2). There is
  also evidence for a morphological-type dependence in the TFR, with
  later types having a steeper relation
  \citep{2006ApJ...653..861M}. By selecting late type galaxies, the
  true TFR in our sample may be steeper than in the
  \citeauthor{2007AJ....134..945P} sample. Larger,
  but less broadly selected, Malmquist corrected TFR samples are
  available \citep[e.g][]{2007ApJS..172..599S,2006ApJ...653..861M}. The narrower selection
  of only late-type spirals in these larger samples matches better the
  colour selection we employ (see Section~\ref{sec:data}). However, we do not
  include any additional morphological pruning (e.g. for peculiar
  galaxies or mergers) common in these larger data sets. We also will
  not include a correction for Malmquist-type biases in our sample
  (see Appendix~\ref{sec:malmquist}). Ultimately, our goal is to look
  for \emph{differences} in the TFR as a function of environment, so
  while the adoption of the \citeauthor{2007AJ....134..945P} TFR for our
  analysis may affect the overall TFR slope recovered, our method is
  self consistent and differences in slope with environment will not
  be adversely affected by this choice of comparison TFR. We have
  checked that this is true by rerunning our analysis using a steeper
  TFR (slope $7.85$, based on \cite{2006ApJ...653..861M}) and find no
  difference in our conclusions (see Section~\ref{sec:differentTFR}).

\section{Data and Sample Selection}\label{sec:data}

The data used for this investigation come from the SDSS DR 8. The SDSS is a photometric and spectroscopic survey mapping approximately $\pi$ steradians of the celestial sphere. Galaxies are imaged in high quality in five bands ($u$, $g$, $r$, $i$ and $z$), which provide reliable magnitude measurements. SDSS galaxy spectra are obtained through a $3''$ diameter fibre. Initially, we select a large sample of $71,933$ galaxies with minimal amount of cuts using the online `CasJobs' catalogue archive server SQL database interface\footnote{http://skyservice.pha.jhu.edu/CasJobs/}. This sample, Sample $0$, was selected according to the following criteria:

(1) We selected galaxies using a colour cut with $u-r<2.22$. The
purpose of this colour cut is to filter galaxies to choose ones with
disc type morphology. The paper by \cite{2001AJ....122.1861S} finds
that SDSS galaxies on a luminosity versus colour diagram have a
bimodal distribution, with early type galaxies (redder) mostly having
$u-r>2.22$ and late type galaxies mostly having
$u-r<2.22$. \cite{2001AJ....122.1861S} determine a dividing colour
between the two populations at $u-r=2.22$ by minimizing the overlap of
two Gaussians that define the two populations. This colour cut allows
us to select mostly spirals, but we are excluding the red-end, dusty
spirals as a consequence. \cite{2004ApJ...600..681B} show that dusty
and bulge-dominated spirals may be found at $u-r>2.22$ so these would
be missing from our analysis. Therefore, our sample primarily consists
of very late type (disc dominated) galaxies without significant
bulges. Moreover, $u-r$ colour is a function of inclination,
  with edge on galaxies appearing about 0.2 mag redder than face on
  galaxies, so this colour cut creates a slight preference for lower
  inclination galaxies \citep{2010MNRAS.404..792M}. An alternative
sample selection criteria one could use for future studies would be to
use the visual classifications from the Galaxy Zoo project
\citep{2008MNRAS.389.1179L}, which could be used to selected a broad
sample of disc type galaxies from very disc-like ones to very bulgy
ones. However, our sample selection of very late type spirals does
help disentangle environmental effects on the TFR related to bulges
(bias could be introduced since there are more bulges in disc galaxies
at higher densities (\citealt{1980ApJ...236..351D},
\citealt{2007MNRAS.382..801M})).

(2) We select galaxies with inclination above $55^\circ$. This inclination cut was made to choose only galaxies that are near edge-on rather than face-on in order to obtain reliable inclination-corrected velocity widths due to the rotation of the disc. At lower inclinations, the error on the line-of-sight velocity becomes comparable to the the velocity. The inclination of a galaxy was estimated from the measured axis ratio ($b/a$) for the galaxy using the formula:
\begin{equation}
\label{eqn:inc}
\sin(i)=\sqrt{\frac{1-(b/a)^2}{1-0.19^2}},
\end{equation}
which accounts for the finite thickness of the disc (a discussion of
this equation is found in \citealt{1984AJ.....89..758H}, and the
particular form of the equation we write here comes from
\cite{2007AJ....134..945P}).  Axis ratios $b/a$ are measured in five
bands ($ugriz$). For each galaxy we average $\sin(i)$ as implied by
the $gri$ bands. We avoid using the $u$ and $z$ bands for inclination
estimates since they are less sensitive to surface brightness
features. We also do not just use a single band because a single band
yields a histogram of the sin inclinations which shows a few narrow
troughs and peaks on top of an otherwise flat distribution most likely
due to the fitting algorithm, and averaging results gives a flattened
histogram (as would be expected due to random inclinations). We select
galaxies with $\sin(i)>0.6$ (corresponding to $>55^\circ$ inclination)
for our sample.

(3) We select galaxies that have emission line measurements from the
MPA-JHU spectroscopic re-analysis. Velocity widths (standard
deviations) are measured simultaneously in all of the Balmer lines, as
described in \cite{2004ApJ...613..898T} and
\cite{2004MNRAS.351.1151B}, where the full details may be found. These
line width measurements are made with a more improved accuracy in
continuum subtraction than the default SDSS spectroscopic pipelines.
 
(4) We only select galaxies that have good photometry with no
important warning flags and accurately measured spectroscopic
redshifts with no warning flags.
 
(5) We make a redshift cut to select galaxies between
$0.005<z<0.085$. The reason for this is that only galaxies in this
redshift range can have accurate projected neighbour densities
($\Sigma$) measured by SDSS. A discussion of $\Sigma$ will follow
shortly in this section.

The data retrieved from SDSS for each galaxy are the following:
spectroscopic redshift $z$; model magnitudes in the five bands
($ugriz$) corrected for Galactic extinction computed following
\cite{1998ApJ...500..525S}; axis ratios $b/a$ in the five bands;
$K_{\rm corr}$ values in the five bands (from the Photoz table in
CasJobs) used for cosmological correction in the calculation of
absolute magnitudes; the standard deviation, $\sigma_{\rm raw}$, of
the Balmer line widths; and total stellar mass estimates of the
galaxies (median estimates were used) derived in
\cite{2003MNRAS.341...33K}, where full details may be found (see also
\citealt{2004MNRAS.351.1151B} and \citealt{2008MNRAS.388..945B}).
photometry of the galaxies.

Absolute magnitudes, $M$, in each band are calculated from the apparent magnitude $m$ and redshift $z$ according to
\begin{equation}
M=m-5\log_{10} (D_{\rm L}/ {\rm pc})-5 - K_{\rm corr}
\end{equation}
where $D_{\rm L}$ is the luminosity distance, which we approximate as:
\begin{equation}
\begin{split}
D_{\rm L}(z) &= \left(1.83\log_{10}(1+z)+0.375\sin(0.67 z)\right) \\ & \quad \times (1+z)\times 4.23\times 10^6 {\rm kpc}.
\end{split}
\end{equation}
The raw line widths $\sigma_{\rm raw}$ (standard deviations) are converted to inclination-corrected FWHMs according to:
\begin{equation}
v_{\rm FWHM}=2.3548\frac{\sigma_{\rm raw}}{\sin i}.
\end{equation}

We ultimately wish to study the TFR as a function of environment, so we need a measure of environmental density around galaxies. \cite{2006MNRAS.373..469B} calculate projected neighbourhood densities, $\Sigma$, for SDSS DR 4 galaxies. To estimate the density, \cite{2006MNRAS.373..469B} determine for each galaxy the $N$th nearest neighbour projected densities:
\begin{equation}
\Sigma_N=\frac{N}{\pi d^2_N},
\end{equation}
where $d_N$ is the projected comoving distance to the $N$th nearest neighbour to a galaxy that is a member of the density defining population (DDP) of galaxies within $\pm \Delta z c =1000$~km~s$^{-1}$. The DDP consists of galaxies with $M_r<-20-1.6(z-0.05)$, that is, it consists of a volume-limited sample (for $z\leq 0.085$). Full details for the calculation of $\Sigma_N$, including corrections for galaxies near the photometric edge, may be found in \cite{2006MNRAS.373..469B}. The best estimate for the projected neighbourhood densities $\Sigma$ is found by \cite{2006MNRAS.373..469B} to be:
\begin{equation}
\log_{10}\Sigma = \frac{1}{2}(\log_{10}\Sigma_4 + \log_{10}\Sigma_5).
\end{equation}
Baldry recalculates $\Sigma$ for SDSS DR 7 galaxies, which are available on his research page\footnote{http://www.astro.ljmu.ac.uk/$\sim$ikb/research/}. We cross-match the spectroscopic IDs of these sources with our Sample $0$ from DR 8. We obtain a new sample, Sample A, of $44,276$ galaxies with density measurements, keeping only galaxies with $v_{\rm FWHM}>50$~km~s$^{-1}$. Sample A was the sample we used in the analysis of Section~\ref{sec:why} in order to investigate at which redshifts might we be able to obtain a TFR with SDSS line widths.

From Section~\ref{sec:why} we have learned that it is best to use a sample of galaxies with $z\geq 0.045$ to reduce aperture biases and construct an SDSS-based TFR. The study by \cite{2005PASP..117..227K} have also found that using the SDSS spectra of the inner parts of galaxies at $z<0.04$ can have strong systematic effects in the estimation of their global properties such as star formation rate, metallicity and reddening. We create Sample B, consisting of $25,698$ galaxies, by eliminating galaxies with $z<0.045$ from Sample A. Sample B is the final sample we construct, which we use for studying the TFR as a function of environment. Galaxies in Sample B have typical magnitudes $-22.5<M_r<-18.5$.

\section{Results}\label{sec:results}

The aperture-biased emission profile modelling we performed in Section~\ref{sec:why} explains why we can expect to obtain a TFR using SDSS line widths measured by a $3''$ fibre. In this section, we demonstrate that we do obtain linear TFRs using the actual data, which makes it possible to study the TFR as a function of environment. 

\begin{figure}
\centering
\includegraphics[width=0.47\textwidth]{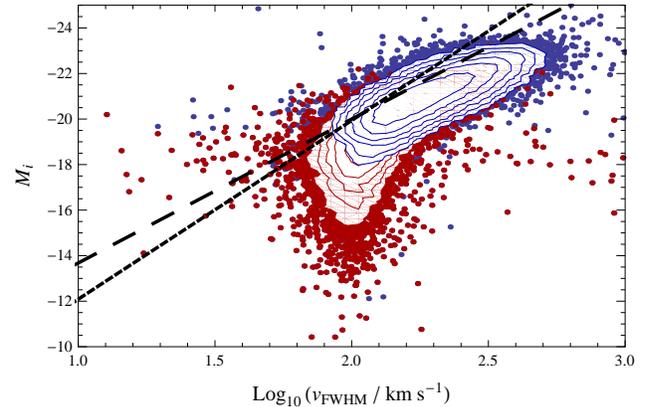}
\caption{Plot of absolute $i$-band magnitude versus inclination-corrected $v_{\rm FWHM}$ for a sample of $71,933$ SDSS galaxies. Galaxies with $z < 0.045$ are in red, and higher redshift galaxies in blue.  Over-saturated points are plotted as log-scale contours. This is a Tully-Fisher plot, except that there is a turn-off due to strong aperture bias at low redshifts $z<0.045$ (red). The distribution shows the same characteristic distribution as predicted by our disc emission models in Figure~\ref{fig:whyTF}. For comparison, the plot also shows the $i$-band TFR as a function of $v_{80}$ from \protect\cite{2007AJ....134..945P} (long-dashed line) and from \protect\cite{2006ApJ...653..861M} (short-dashed line). The average $z$ for the data clump of galaxies that turn off from a linear trend at $M_i>-17$~mag is $z=0.011$.}
\label{fig:MiAll}
\end{figure}

Firstly, we plot absolute magnitude in the $i$-band, $M_i$, versus $v_{\rm FWHM}$ for Sample $0$, presented in Figure~\ref{fig:MiAll}. This sample contains galaxies with redshifts $0.005<z<0.085$, so we expect to see strong aperture effects from the low-redshift galaxies. The purpose of Figure~\ref{fig:MiAll} is to see where the galaxies lie and compare its general shape with the plots of Figure~\ref{fig:whyTF} which shows the predicted shape of the TFR from our simple disc models. We do see the low-$z$ (hence low luminosity) turn-off of galaxies, and we are justified in excluding galaxies with $z<0.045$ in Sample B. There is also some degree of turn-off expected for low-luminosity galaxies with $v_{\rm circ}<90$~km~s$^{-1}$ due to physical reasons \citep{2000ApJ...533L..99M}. This turn-off is due to the fact that these low luminosity galaxies are very rich in gas, and if the gas mass in the disc is added to the stellar mass, then the turn-off can be rectified and these galaxies may be returned to the TFR trend line. However, we excluded these low-luminosity galaxies in Sample B when we made our redshift cut, and therefore do not have to worry about correcting for this effect.

\begin{figure*}
\centering
\includegraphics[width=\textwidth]{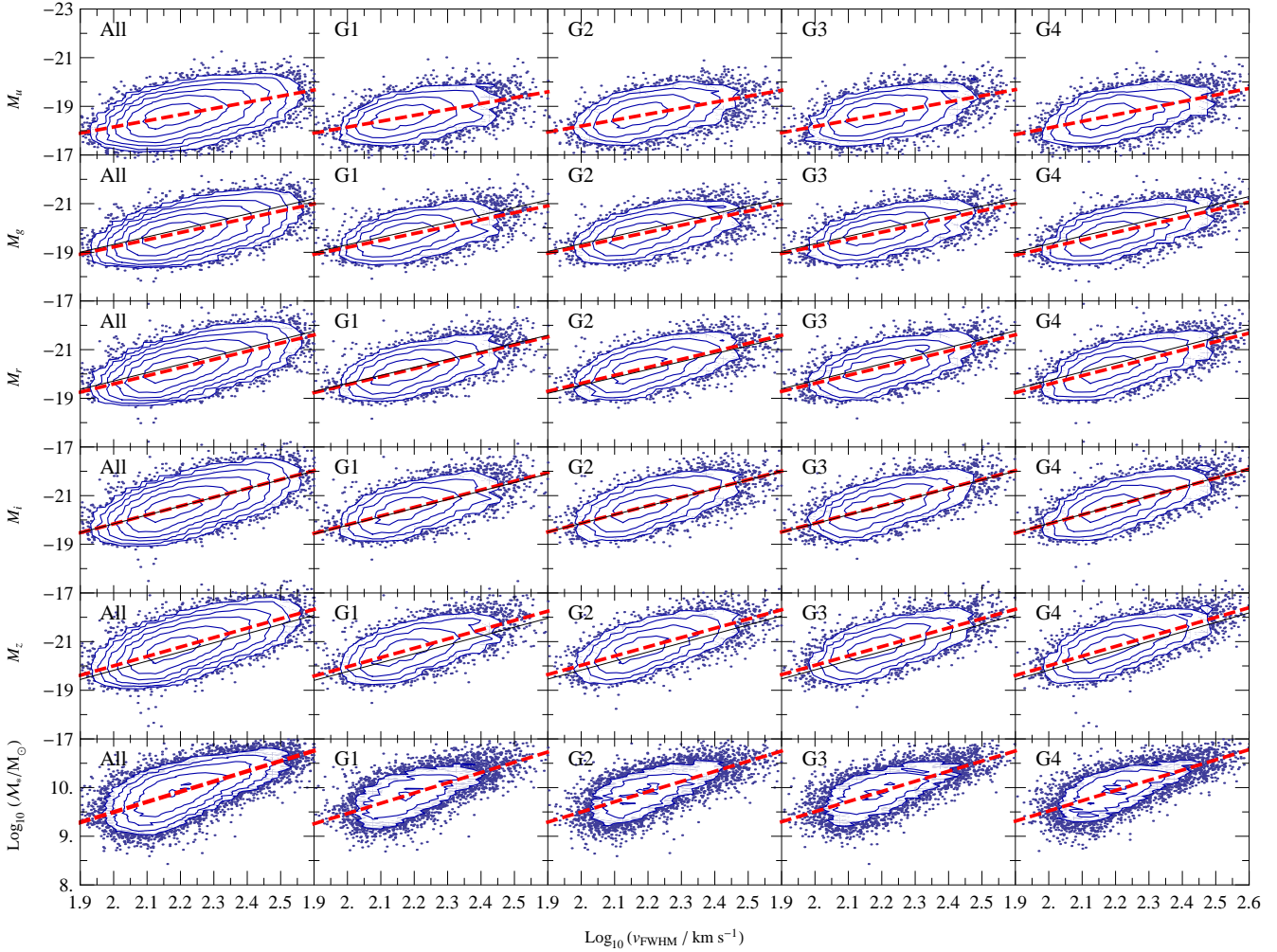}
\caption{TFRs in five bands and smTFRs for SDSS all galaxies (Sample B) and galaxies divided into groups by density. Over-saturated points are plotted as log-scale contours. G$1$ represents the group of galaxies with the lowest $25$ per cent of environmental densities and G$4$ represents the highest $25$ per cent. The slopes and scatter in each band appear mostly independent of environmental density, with slight increase of slope with higher density in the bluer bands. Thick, red, dashed lines show the best-fit linear models. Thin black lines show the $griz$ band relations expected from simulations (see Section~\ref{sec:environment}) and assuming the TFRs of \protect\cite{2007AJ....134..945P}.}
\label{fig:TFRall}
\end{figure*}

We plot the SDSS-based TFRs in five bands ($ugriz$) for Sample B in
Figure~\ref{fig:TFRall}. We obtain linear trends with large scatter
($\sim0.5$~mag) due to the aperture effects as predicted in
Section~\ref{sec:why}.  The estimated additional scatter by aperture
bias is $\sim0.3$--$0.4$~mag, comparable to the $0.4$~mag intrinsic
TFR scatter found in \cite{2007AJ....134..945P}.  Linear models are
fitted to data in the five bands, with equations listed in
Table~\ref{tbl:TFR}.  We perform forward fits, and have not included
errors on individual points in fitting the relation, an approach
justified due to the uniformity of the SDSS dataset across such a
large sample of galaxies (i.e. the errors on each point are
effectively the same, so including them does not affect the result,
but does greatly increase the processing complexity of the problem).
We see steepening in the TFR from $-2.53\pm0.06$~mag~$(\log_{10}~{\rm
  km}~{\rm s}^{-1})^{-1}$ in the $u$ band to
$-3.87\pm0.06$~mag~$(\log_{10}~{\rm km}~{\rm s}^{-1})^{-1}$ in the $z$
bands (it is expected for the TFR to steepen in redder bands
\citep{2000ApJ...533..744T, 2001ApJ...563..694V, 2002A&A...396..431K,
  2007ApJ...671..203C, 2007AJ....134..945P,
  2008AJ....135.1738M}). There is no significant difference in the
scatter between the different bands, since the aperture effects
significantly contribute to the scatter. In an ordinary TFR, one may
expect less scatter in the redder bands \citep{1979ApJ...229....1A,
  2000ApJ...533..744T, 2008AJ....135.1738M}, where the mass to
luminosity ratio of galaxies is more homogeneous since the redder
bands are less sensitive to star formation and more sensitive to the
older population of stars which dominate the mass of the galaxy
(although \cite{2007AJ....134..945P} find the same intrinsic scatter
in the $g$, $r$, $i$ and $z$ bands, possibly because of the fairly
small sample size). In Figure~\ref{fig:TFRall} we also plot the smTFR
for the sample, and the best linear fit to describe
$\log_{10}(\mathcal{M}_*/M_\odot)$ versus $\log_{10}(v_{\rm FWHM} /
{\rm km}~{\rm s}^{-1})$ is found in Table~\ref{tbl:TFR}.

For simplicity we have not included an inclination dependent extinction correction
  in our analysis. Extinction due to dust in the disks of spiral
  galaxies is a strong function of viewing angle
  \citep{2010MNRAS.404..792M,
    2009ApJ...691..394M}. \cite{2010MNRAS.404..792M} find extinction
  of a galaxy viewed edge on is 0.7, 0.6, 0.5, and 0.4 mag greater
  than for a face on view for the Sloan $ugri$ passbands,
  respectively.  \citeauthor{2009ApJ...691..394M} show that the
  extinction is a linear function of $\log_{10}{b/a}$, where $b/a$ is
  the observed axis ratio of the galaxy. For galaxies with $55\degr < i
  < 90\degr$, this implies an extinction range in e.g. $i$-band of
  $0.13$ to $0.4$ mag. While this range is significant, it is still smaller
  than the overall scatter of $\sim 2.5$ magnitude. Future analysis of
  SDSS data in this way would benefit from including an inclination
  dependent correction, however, since the distribution of
  inclinations should not correlate with environment, the omission
  here does not affect our conclusions.

\subsection{TFR as function of environment}\label{sec:environment}

To investigate the environmental dependence of the TFR, we divide Sample B into four equal-sized groups based on projected density quartiles. The minimum, maximum, and quartile values of $\log_{10}(\Sigma/{\rm Mpc}^{-2})$ are:
\begin{equation} 
\{\min,Q_1,Q_2,Q_3,\max\} = \{-2.06,-0.74,-0.43,-0.06,1.96\}. 
\end{equation}
Call the four groups G$1$, G$2$, G$3$ and G$4$, with G$1$ (see Table~\ref{tbl:Gdef}) corresponding to the galaxies with the least dense environments typical for galaxies in void-like regions and G$4$ corresponding to galaxies in the most dense environments typical to cluster galaxies and galaxies at the peripheries of clusters. Here, we investigate whether there is a significant difference in the TFRs of galaxies belonging to these four different groups.

We plot the SDSS fibre-based TFR as a function of $v_{\rm FWHM}$ for each group in each of the bands in Figure~\ref{fig:TFRall}. The best-fit linear models (forward fits) for the TFRs are listed in Table~\ref{tbl:TFR}. In Figure~\ref{fig:TFRall} we also plot the SDSS-based smTFRs for each of the four density groups, with best fits listed in Table~\ref{tbl:TFR}. The $95$ per cent confidence (statistical) errors on the slopes and intercepts from $100$ bootstrap realizations are included in the tables in parentheses (i.e., to obtain errors, for each realization the data set is re-sampled with replacement and fit with a linear relation to obtain a probability distribution for the slope and intercept). The slopes of these TFRs as a function of $v_{\rm FWHM}$ are approximately half the value of the slopes of real TFRs as a function of $v_{\rm circ}$. This is attributed to the large scatter and flattened slope in the relation between $v_{\rm circ}$ and $v_{\rm FWHM}$ due to aperture bias. We show shortly in this section with the aid of simulations that these flattened fibre-based TFR slopes are expected if we assume that the relation between magnitude and $v_{\rm circ}$ are the TFRs of \cite{2007AJ....134..945P}. 

\begin{table}
\centering
\caption{Definitions of density groups}\label{tbl:Gdef}
\begin{tabular}{@{}ll@{}}
\hline\hline
\noalign{\smallskip}
group & $\log_{10}(\Sigma/{\rm Mpc}^{-2})$ \\[2pt]
\hline 
\noalign{\smallskip}
G$1$ & $[-2.06,-0.74]$ \\
G$2$ & $[-0.74,-0.43]$ \\
G$3$ & $[-0.43,-0.06]$ \\
G$4$ & $[-0.06,1.96]$ \\
\hline
\end{tabular}
\end{table}

\begin{table*}
\centering
\caption{TFRs in five bands and smTFRs for the entire sample and environmental density groups (as function of $v_{\rm FWHM}$)}\label{tbl:TFR}
\small
\begin{tabular}{@{}lccc@{}}
\hline\hline
\noalign{\smallskip}
     & $u$ & $g$ & $r$ \\
All  & $-13.08(0.14) - 2.53(0.06) v$ & $-13.31(0.13) - 2.95(0.06) v$ & $-12.88(0.13) - 3.36(0.06) v$ \\
G$1$ & $-13.30(0.29) - 2.42(0.13) v$ & $-13.51(0.23) - 2.85(0.11) v$ & $-13.01(0.25) - 3.28(0.12) v$ \\
G$2$ & $-13.29(0.27) - 2.45(0.12) v$ & $-13.49(0.23) - 2.88(0.11) v$ & $-13.04(0.24) - 3.29(0.11) v$ \\
G$3$ & $-13.11(0.29) - 2.53(0.13) v$ & $-13.39(0.23) - 2.93(0.11) v$ & $-12.96(0.31) - 3.33(0.13) v$ \\
G$4$ & $-12.70(0.30) - 2.71(0.13) v$ & $-12.98(0.30) - 3.11(0.14) v$ & $-12.60(0.28) - 3.49(0.14) v$ \\[2pt]
\hline 
     & $i$ & $z$ & sm \\
All  & $-12.56(0.14) - 3.64(0.06) v$ & $-12.26(0.14) - 3.87(0.06) v$ & $5.28(0.08) + 2.11(0.04) v$ \\
G$1$ & $-12.66(0.26) - 3.58(0.13) v$ & $-12.35(0.28) - 3.81(0.13) v$ & $5.23(0.13) + 2.12(0.06) v$ \\
G$2$ & $-12.71(0.25) - 3.58(0.11) v$ & $-12.42(0.28) - 3.80(0.13) v$ & $5.31(0.15) + 2.09(0.07) v$ \\
G$3$ & $-12.65(0.32) - 3.60(0.14) v$ & $-12.36(0.32) - 3.84(0.15) v$ & $5.33(0.15) + 2.09(0.07) v$ \\
G$4$ & $-12.31(0.34) - 3.76(0.15) v$ & $-12.05(0.23) - 3.98(0.15) v$ & $5.31(0.18) + 2.10(0.08) v$ \\
\hline	
\end{tabular}
\end{table*}

\begin{figure}
\centering
\includegraphics[width=0.47\textwidth]{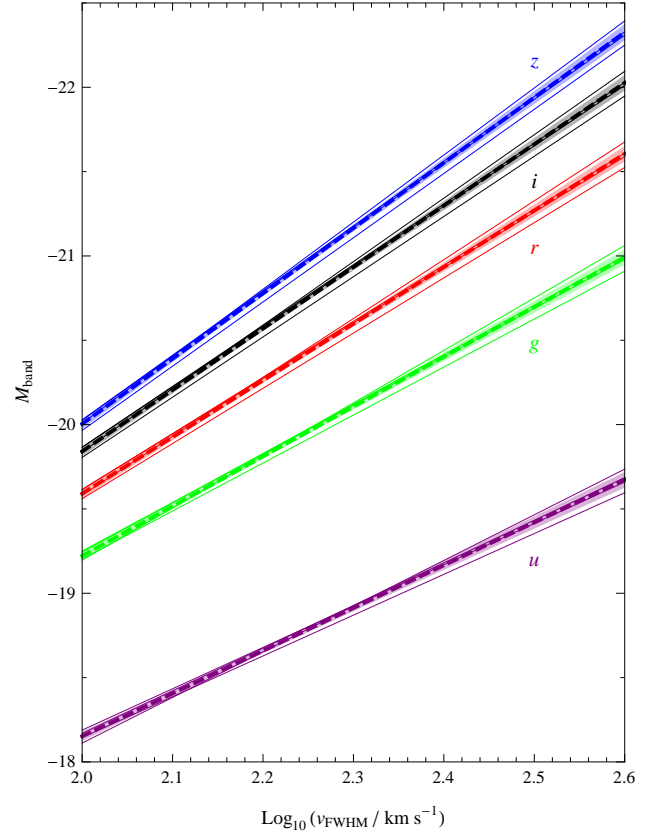}
\caption{TFR in the five bands for all SDSS galaxies in Sample B as well as the different density groups. The colours violet, green, red, black and blue correspond to $u$, $g$, $r$, $i$ and $z$ bands respectively. The thick dashed lines show the TFR for all the galaxies in Sample B. The shaded areas represent the $99.99$ per cent mean prediction bands for the relationship. The thin lines show the TFR for the different density groups for the five bands. The TFRs for the least dense and most dense groups lie outside the shaded confidence region in all five bands.}
\label{fig:main}
\end{figure}

First, we notice that the TFRs for the four environmental density groups are similar to each other for each of the five bands. There does appear to be a small amount of steepening of the relations with denser environments. The TFR steepens $11.8$, $9.2$, $6.5$, $5.2$ and $4.3$ per cent in the $u$, $g$, $r$, $i$ and $z$ bands respectively from the least dense to most dense groups. The relative steepening is larger in bluer bands. In Figure~\ref{fig:main} we plot the TFR in each of the five bands for the entire sample, showing (statistical) $99.99$ per cent mean prediction bands for the relation, and plot the TFRs derived for each density group. The TFRs for groups G$2$ and G$3$ lie within the confidence region, but the TFRs for groups G$1$ and G$4$ lie outside. At this point, we devise several tests to investigate whether this steepening is significant and whether it is due to the environment.

To test the significance of the differences in slope, we randomly permute the densities of the galaxies and repeat dividing the galaxies by density quartiles and fitting each group with a linear model for $2500$ realizations. For each realization, we compute the steepening ratio between the highest and lowest sloped groups. From the distribution of the ratios, we find that the steepening in the $u$ band is significant at the $0.01$ probability level and the steepening in the $g$ band is significant at the $0.02$ probability level. Steepening in the other bands are not significant at the $0.05$ probability level. TFRs assuming that the lines are parallel in each of the four density groups are presented in Table~\ref{tbl:para}. The variables $z_1$, $z_2$, and $z_3$ are indicator variables, such that $z_i=1$ for a galaxy if it is in group G$i$, else $z_i=0$. In the $u$ and $g$ bands, where parallelism is rejected at the $3\sigma$ level, there is no coherent pattern in the $y$-intercept between the density groups assuming parallelism. In the $r$, $i$, and $z$ bands the TFRs marginally brighten with increasing density, assuming the relations are parallel across the density groups. The smTFR shows marginal increase in stellar mass at fixed velocity width with increased density.

\begin{table}
\centering
\caption{TFRs for density groups assuming parallelism}\label{tbl:para}
\begin{tabular}{@{}cc@{}}
\hline\hline
\noalign{\smallskip}
band & TFR \\[2pt]
\hline 
\noalign{\smallskip}
$u$ & $-13.09-2.53 v+0.028 z_1-0.022 z_2-0.019 z_3$ \\
$g$ & $-13.34-2.95 v+0.048 z_1-0.0055 z_2-0.0085 z_3$ \\
$r$ & $-12.91-3.35 v+0.067 z_1+0.0088 z_2+0.0023 z_3$ \\
$i$ & $-12.59-3.64 v+0.073 z_1+0.012 z_2+0.0051 z_3$ \\
$z$ & $-12.30-3.87 v+0.077 z_1+0.018 z_2+0.0096 z_3$ \\
stellar & $5.32+2.10 v-0.052 z_1-0.024 z_2-0.019 z_3$ \\
\hline
\end{tabular}
\end{table}

\begin{figure}
\centering
\includegraphics[width=0.47\textwidth]{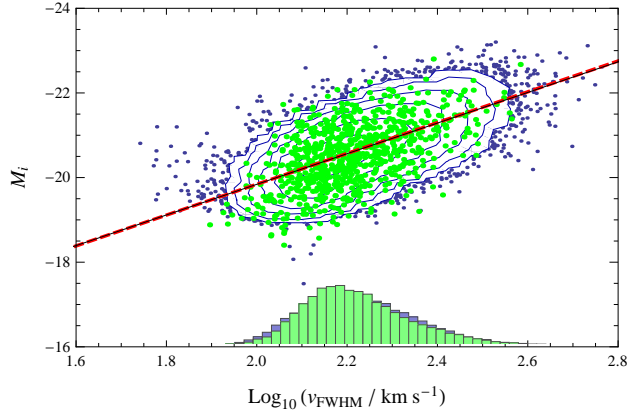}
\caption{The observed $i$-band data (blue) with the best fit SDSS-based TFR (thick, red dashed line) compared to simulated data (green; $800$ data points are shown for clarity) and its best-fit linear relation (thin, black line) which assumes the TFR of \protect\cite{2007AJ....134..945P} and the galaxy size distribution of our sample. There is very good agreement between the two, despite the simplifying assumptions we have made.
A histogram of the $v_{\rm FWHM}$ for the real and simulated data is also shown, showing close agreement, but slightly more dispersion in the real data than the simplified model.}
\label{fig:comp}
\end{figure}

\begin{figure}
\centering
\includegraphics[width=0.47\textwidth]{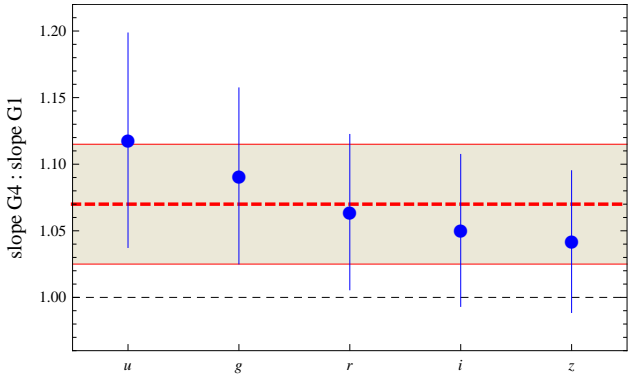}
\caption{The simulated population of galaxies using the observed parameter distribution predict a steepening of $7\pm4$ (red dashed line and shaded region) per cent in the slope of the SDSS fibre-based TFR from the lowest to the highest density group in all magnitude bands, due to small variations in the size distributions of the different density groups (assuming no trends with density). The observed data agrees, within error, with this prediction. Without aperture biases, no steepening would be expected (thin, black dashed line) if there were no environmental trends with density.}
\label{fig:steep}
\end{figure}

A difference in the linear relations between magnitude and $v_{\rm FWHM}$ for two groups of data may not necessarily imply that the TFR is not the same for the two groups. As we learned from our simulated population and Figure~\ref{fig:whyTF}, the slope is affected by the size distribution of the galaxies. Differences in size distributions in the two groups may therefore cause the difference in slope. Thus we create synthetic samples of galaxies (same sample size as the observed sample) for each of the four density groups, with model parameters drawn from the observed distribution of sizes (and again we assume the local TFR to estimate a distribution for disc velocities). The purpose of the simulations is to see whether they predict the same slopes for the relation between magnitude and $v_{\rm FWHM}$ as observed for the four density groups, and also to obtain errors on the slopes. The distributions of parameters for galaxies in each of the four density groups are listed in Table~\ref{tbl:gpars}, where it can be seen that there are some small differences which may affect the slope. The distribution of sizes at a fixed stellar mass is independent of environment in general \citep{2004MNRAS.353..713K}, however the slope of $v_{\rm circ}$ versus $v_{\rm FWHM}$ may be sensitive to small discrepancies in sizes. We fit the relation between $v_{\rm circ}$ and $v_{\rm FWHM}$ along with errors predicted from $30$ synthetic sample realizations. The results are shown in Table~\ref{tbl:vv}, which indicates slight steepening in $v_{\rm circ}$ versus $v_{\rm FWHM}$ with density (and tight errors) and says that the observed SDSS-based TFR slopes should be around $1/2$ the values of the real slopes. Next, we convert the $v_{\rm circ}$ to $griz$ magnitudes using the TFRs found by \cite{2007AJ....134..945P} and add random Gaussian scatter of $0.4$~mag (\cite{2007AJ....134..945P} find for their sample of $162$ galaxies that the TFR residuals are approximately Gaussian). We use $v_{80}=v(3.03h)=0.80v_{\rm circ}$ in the conversion of $v_{\rm circ}$ to magnitude for our exponential-profile disc models. Doing so, we have created synthetic magnitude versus $v_{\rm FWHM}$ plots just like the observed data and can use it to predict differences in the relation in the four density groups as well as better estimates of error in the linear fit parameters than with bootstrapping (see Figure~\ref{fig:comp}). The SDSS-based TFRs for our synthetic population are presented in Table~\ref{tbl:synTFR} and plotted along with the real data in Figure~\ref{fig:TFRall} as black, solid lines in the $griz$ bands. The predicted and observed slopes and intercepts agree remarkably well given the simplifying assumptions we have made. From the simulations we also obtain a good sense of what the errors on the parameters of the fits to the real data should be, and find that they are comparable to the bootstrap errors we previously found. Notably, we predict steepening on the order of $7\pm4$ per cent of the SDSS-based TFRs from the lowest to the highest density group in all the bands due to size distribution differences in the groups (see Figure~\ref{fig:steep}).

The agreement between our simulated data and observations suggests
that there is no environmental effect on our TFRs. We can ask now how
sensitive is our method, that is, how much would the slope of the TFR
as a function of $v_{\rm circ}$ have to change so that our synthetic
populations would predict an SDSS-based TFR that lies outside the
$95$~per~cent confidence interval for the observed SDSS-based TFRs. To
answer this, we perturb the slopes of the TFRs of
\cite{2007AJ....134..945P}, which we have used in our simulations,
about $\log_{10}(v_{\rm circ}/{\rm km}~{\rm s}^{-1})=2.2$ and repeat
the simulations to predict the slopes of the observed SDSS-based
TFRs. We find that we need to perturb the slopes by $\lesssim
3$~per~cent so that the predicted fibre-based TFRs lie outside the
$95$~per~cent confidence intervals of the observed relations in
Table~\ref{tbl:TFR}. Similarly, we find that we can perturb the
intercept of the TFR by $\lesssim 2$~per~cent so that the predicted
fibre-based TFRs lie outside the $95$~per~cent confidence intervals of
the observed relations. Therefore, we can say that if there are
environmental effects on the TFR in our sample, then it would be very
subtle, having less than $3$~per~cent on TFR parameters. This
$3$~per~cent variation is comparable to the errors on the observed
TFRs in the first place.

\subsubsection{Another test for environmental effects on the TFR across bands}\label{sec:slopeRatio}

Another test we can perform to show that the steepening in the TFRs from the least to most dense groups is not due to environmental effects is to look at the observed ratio of steepening across different bands and compare it across bands.
An important realization is that given two groups of $\log_{10}v_{\rm circ}$ versus $\log_{10}v_{\rm FWHM}$ data which may have slightly different slopes $s_1$ and $s_2$, if the $y$-axis is converted to magnitudes using the real TFR in order to obtain magnitude versus $\log_{10}v_{\rm FWHM}$ plots (such as the observed data), then the ratio of the new slopes $s^\prime_1,s^\prime_2$ of the two groups should be the \textit{same in any magnitude band}. This is because if the TFR is of the form $M=a \log_{10}(v) + b$ then:
\begin{eqnarray}
\frac{s_1}{s_2}
&=& \frac{\Delta_1 (\log_{10}v_{\rm circ}) / \Delta_1 (\log_{10}v_{\rm FWHM})}{\Delta_2 (\log_{10}v_{\rm circ})/\Delta_2 (\log_{10}v_{\rm FWHM})} \\
&=& \frac{a\Delta_1 (\log_{10}v_{\rm circ}) / \Delta_1 (\log_{10}v_{\rm FWHM})}{a\Delta_2 (\log_{10}v_{\rm circ})/\Delta_2 (\log_{10}v_{\rm FWHM})} \\
&=& \frac{\Delta_1 M / \Delta_1 (\log_{10}v_{\rm FWHM})}{\Delta_2 M/\Delta_2 (\log_{10}v_{\rm FWHM})} \\
&=&\frac{s^\prime_1}{s^\prime_2}.
\end{eqnarray}
This gives the most powerful test to see whether the TFR alters with density because we can simply compare the ratios of steepening across bands and do not have to rely on the simple model, or a particular assumed ``real'' TFR, or trying to correct $v_{\rm FWHM}$ to some $v_{\rm circ}$. The simulated data though is necessary to give us an estimate on the error/variation allowed on the slope ratios for the TFR to be parallel in all the density subgroups (but the magnitude of the errors is not expected to vary significantly with model parameters). We have observed a slope increase in the actual data of $11.8$, $9.2$, $6.5$, $5.2$ and $4.3$ per cent in the $ugriz$ bands, while the simulated data predict a steepening of $7\pm 4$ per cent (see Table~\ref{tbl:nodep}). This is consistent with no dependence of the TFR on environment. A similar test for the intercept of the TFR finds no dependence on environment (see Table~\ref{tbl:nodep}).

In summary, the simulations show that steepening is expected with
environment in the SDSS-based TFRs due to the small differences in
size distributions for the different density groups, and most
importantly that the observed slopes and intercepts of the SDSS-based
TFRs are consistent with the TFRs of \cite{2007AJ....134..945P} at the
$95$ per cent level (compare the red and black lines in
Figure~\ref{fig:TFRall}).

\begin{table}
\centering
\caption{Gaussian probability distributions for observed parameters in density groups}\label{tbl:gpars}
\begin{tabular}{@{}ccccc@{}}
\hline\hline
\noalign{\smallskip}
 & \multicolumn{2}{c}{$\log_{10}(h/\rm{arcsec})$} & \multicolumn{2}{c}{$\log_{10} (v_{\rm circ} / \rm{km}~\rm{s}^{-1})$} \\
group & $\mu$ & $\sigma$ & $\mu$ & $\sigma$ \\[2pt]
\hline 
\noalign{\smallskip}
G$1$ & $0.521$ & $0.178$ & $2.08$ & $0.101$ \\
G$2$ & $0.524$ & $0.173$ & $2.10$ & $0.102$ \\
G$3$ & $0.522$ & $0.171$ & $2.10$ & $0.103$ \\
G$4$ & $0.521$ & $0.164$ & $2.10$ & $0.105$ \\
\hline
\end{tabular}
\end{table}

\begin{table}
\centering
\caption{$v_{\rm circ}$ versus $v_{\rm FWHM}$ from simulations}\label{tbl:vv}
\begin{tabular}{@{}cc@{}}
\hline\hline
\noalign{\smallskip}
group & $\log_{10}(v_{\rm circ})=$ \\[2pt]
\hline 
\noalign{\smallskip}
All & $0.573(0.008)\log_{10}(v_{\rm FWHM}) + 0.923(0.017)$ \\
G$1$ & $0.556(0.012)\log_{10}(v_{\rm FWHM}) + 0.953(0.025)$ \\
G$2$ & $0.563(0.017)\log_{10}(v_{\rm FWHM}) + 0.946(0.035)$ \\
G$3$ & $0.569(0.011)\log_{10}(v_{\rm FWHM}) + 0.936(0.023)$ \\
G$4$ & $0.595(0.015)\log_{10}(v_{\rm FWHM}) + 0.88(0.031)$ \\
\hline
\end{tabular}
\end{table}

\begin{table*}
\centering
\caption{Expected TFRs (as function of $v_{\rm FWHM}$) from synthetic populations assuming simple emission model}\label{tbl:synTFR}
\small
\begin{tabular}{@{}lcccc@{}}
\hline\hline
\noalign{\smallskip}
     & $g$ & $r$ & $i$ & $z$\\
All  & $-13.06(0.09) - 3.13(0.04) v$ & $-12.89(0.12) - 3.41(0.06) v$ & $-12.60(0.14) - 3.62(0.07) v$ & $-12.28(0.15) - 3.77(0.08) v$ \\
G$1$ & $-13.19(0.21) - 3.06(0.10) v$ & $-13.01(0.22) - 3.30(0.11) v$ & $-12.74(0.22) - 3.52(0.11) v$ & $-12.47(0.23) - 3.65(0.11) v$ \\
G$2$ & $-13.15(0.22) - 3.10(0.11) v$ & $-13.01(0.27) - 3.26(0.13) v$ & $-12.74(0.32) - 3.56(0.16) v$ & $-12.41(0.31) - 3.71(0.15) v$ \\
G$3$ & $-13.15(0.21) - 3.10(0.10) v$ & $-12.93(0.18) - 3.40(0.09) v$ & $-12.67(0.19) - 3.59(0.09) v$ & $-12.36(0.24) - 3.74(0.11) v$ \\
G$4$ & $-12.84(0.16) - 3.25(0.07) v$ & $-12.65(0.22) - 3.54(0.11) v$ & $-12.33(0.28) - 3.76(0.13) v$ & $-11.99(0.24) - 3.92(0.12) v$ \\[2pt]
\hline	
\end{tabular}
\end{table*}

\begin{table}
\centering
\caption{Steepening due to differences in size and velocity distributions in density groups}\label{tbl:nodep}
\begin{tabular}{@{}ccc@{}}
\hline\hline
\noalign{\smallskip}
 & slope G$4$ / slope G$1$ & intercept G$1$ / intercept G$4$ \\[2pt]
\hline 
\noalign{\smallskip}
predicted & $1.07\pm0.04$ & $1.03\pm 0.02$ \\
obs. $u$ & $1.118\pm0.08$ & $1.047\pm 0.03$ \\
obs. $g$ & $1.092\pm0.07$ & $1.041\pm 0.03$ \\
obs. $r$ & $1.065\pm0.06$ & $1.032\pm 0.03$ \\
obs. $i$ & $1.052\pm0.06$ & $1.027\pm 0.04$ \\
obs. $z$ & $1.043\pm0.05$ & $1.024\pm 0.03$ \\
\hline
\end{tabular}
\end{table}

\subsection{Assuming a Different TFR in the Analysis}\label{sec:differentTFR}

 We have already discussed the limitations of the choice of 
 the \cite{2007AJ....134..945P} TFR (Section~\ref{sec:why2}).
 Here we investigate the effects of using a different choice 
 for the TFR for the analysis. We pick a steeper $i$-band TFR
 of $M_i = -7.85\log_{10}v_{\rm 80}-4.00$. This TFR was 
 obtained by increasing the slope of the \cite{2007AJ....134..945P}
 TFR to $-7.85$ about $\log_{10}v_{\rm 80}=2.2$. This steeper slope of 
 $7.85$ is based on the $I$-band TFR for late-type spiral (Sc)
 galaxies of \cite{2006ApJ...653..861M} (the correction between 
 $I$-band and SDSS $i$-band is small \citep{2006A&A...460..339J}).
 
We repeat all of the procedures with this new TFR, 
 including estimating the distribution of $v_{\rm circ}$ from 
 the  SDSS magnitudes. This analysis is not equivalent to 
 assuming a fixed inferred distribution of galaxy rotation 
 profiles (hence $v_{\rm circ}$ and $v_{\rm FWHM}$) and 
 investigating the effect of altering the TFR on the observed 
 SDSS $v_{\rm FWHM}$-based TFR (which would predict an increase in the slope
 of the observed SDSS-based). Re-inferring 
 the distribution of $v_{\rm circ}$ from the SDSS magnitudes 
 using a steeper TFR leads to a narrower distribution of 
 $v_{\rm circ}$,  which can flatten the slope of the correlation 
 between $v_{\rm circ}$ and $v_{\rm FWHM}$. With this steeper 
 slope, the predicted SDSS-observed $v_{\rm FWHM}$-based TFRs for the density groups
 G$1$, G$2$, G$3$, G$4$ are:
 $M_i = -13.01-3.30 v_{\rm FWHM}$,
 $M_i = -13.03-3.31 v_{\rm FWHM}$,
 $M_i = -12.82-3.42 v_{\rm FWHM}$,
 $M_i = -12.63-3.52 v_{\rm FWHM}$ respectively.
 The size of the errors in the slope and intercept are similar 
 to the previous simulations. The predicted slopes are slightly 
 shallower than observed in the SDSS data. However, the simulations 
 still predict steepening of $1.07$ for slope G$4$ / slope G$1$.

 A steeper TFR does not necessarily mean that the predicted
 SDSS $v_{\rm FWHM}$-based TFR will be steeper, if one has to infer the distribution 
 of $v_{\rm circ}$ from the SDSS data. This is because the correlation
 between $v_{\rm circ}$ and $v_{\rm FWHM}$ may flatten if there is 
 a narrower distribution of $v_{\rm circ}$. This slope may be 
 changed/strengthened if we change the intercept of the assumed-TFR
 as well. Exploring the parameter space further, we find that a steep 
 (slope $7.85$) TFR of $M_i = -7.85\log_{10}v_{\rm 80}-4.25$ is consistent
 with the slopes of the SDSS $v_{\rm FWHM}$-based TFRs. This TFR is equivalent to 
 changing the slope of the \cite{2007AJ....134..945P} TFR to $-7.85$ 
 about $\log_{10}v_{\rm 80}=2.0$.

 It appears that the right choice of two different slopes and intercepts 
 may conspire in certain cases to give the same predicted SDSS $v_{\rm FWHM}$-based TFR. 
 However, there is another prediction made by the simulations, namely 
 the distribution of the observed $v_{\rm FWHM}$. A steeper TFR predicts 
 a narrower distribution of $v_{\rm circ}$ and  consequently 
 $v_{\rm FWHM}$. So a useful check to make sure there is no conspiracy is 
 to investigate whether there are significant differences  in the 
 distribution of the observed distribution of $v_{\rm FWHM}$ across 
 density groups (these differences we find, would be much larger than those
 caused by the small differences in size distribution across density groups). 
 We find none: the mean and standard deviations  of $\log_{10}v_{\rm FWHM}$ 
 are $2.21$ and $0.12$ in all density groups  (with variation $<1$~percent). 
 Based on our simulations, a TFR slope of  $6.32$ (based on 
 \cite{2007AJ....134..945P}) and a steeper TFR slope of $7.85$ (based on 
 \cite{2006ApJ...653..861M}) which predict the same  SDSS $v_{\rm FWHM}$-based TFR, would 
 predict differences in mean and standard deviation in $v_{\rm FWHM}$ at 
 the level of $8$~per cent.

\section{Discussion and Conclusions}\label{sec:disc}

We have demonstrated that we can use SDSS-measured line widths, which suffer from $3''$ fibre aperture bias, to construct TFRs for galaxies at $z>0.045$, where enough of the galaxy fits inside the fibre so that we can recover its global properties. Hence, the SDSS database can be very useful for studying the TFR on a large sample of galaxies, and one of the best applications is to look at the effects of environmental density on the TFR. We did so, and found the following:

\begin{itemize}
\item We divided our sample of galaxies into four groups based on density, and found slight steepening of the SDSS-based TFR slope with density in each of the five bands ($ugriz$), but this was attributed (using simulations) to be a consequence of aperture biases affecting the density groups, which had small variations in size distributions. We conclude that there is \textit{no strong or statistically significant environmental effect} at the $95$ per cent level on the TFR in our sample of $25,698$ galaxies and the data is \textit{consistent with the TFR of \cite{2007AJ....134..945P}}. The environments we investigated ranged from void-like regions, with typical $\Sigma$ of $0.05$~Mpc$^{-2}$, to centres of galaxy clusters, with typical $\Sigma$ of $20$~Mpc$^{-2}$. 
Our method is sensitive enough to allow us to conclude that if there is an effect on the TFR due to environment, the change in the TFR slope or intercept is less than $3$~per~cent.

\item Steepening in the TFR is seen on the order of $4$ to $11$ per cent with increasing density, with the relative steepening being larger in the bluer bands. Our simulated populations of model galaxies suggest a steepening of $7\pm 4$ per cent across all bands, accounting for the size change in the groups. One concern, however, may be that perhaps some of the steepening is due to contamination by early type galaxies in our sample, which can have a biasing effect since they tend to be found in denser environments (\citealt{1980ApJ...236..351D}, \citealt{2007MNRAS.382..801M}) and we had only made a simple colour cut to select disc type morphology and may have included some of the less-red ellipticals in our sample. However, repeating the analysis with a stricter colour cut of $u-r<2.00$, we witness the same degree of increase in the TFRs with increasing density ($11.8$, $9.1$, $6.5$, $5.2$ and $4.3$ per cent in the $ugriz$ bands). In addition, there is no significant difference in the concentration indices $c_i=R_{90}/R_{50}$ of the galaxies in the four density groups ($R_{90}$ and $R_{50}$ are the radii enclosing $90$ per cent and $50$ per cent of the $i$-band Petrosian flux). The concentration index is a measure of morphology -- the significance of the bulge -- for galaxies, with $c_i>2.6$ being typical for early-type galaxies and $c_i<2.6$ being typical for late-type galaxies \citep{2001AJ....122.1861S}. Our colour-cut has eliminated the disc galaxies with significant bulges, which is why we do not see a bias in the distribution of $c_i$ across density groups. In addition, we still witness the steepening with density if we make a concentration index cut to select only galaxies with $c_i<2.6$ (most galaxies in our sample have $c_i<2.6$ to begin with). 
Since our sample primarily contains late disc type galaxies, we are not subject to bias from environmental effects related to the bulges of galaxies (there tend to be more bulges in disc galaxies in higher density environments). We would not detect in our study a trend in the TFR with environment which is really a trend with morphological type, as in the results of \cite{2006ApJ...653..861M}.

\item The environmental dependence of the TFR for late-type spirals is
  small (if any), meaning that the intrinsic properties of a galaxy
  play a principal role in driving the evolution of disc type
  galaxies. Our finding agrees with the earlier work by
  \cite{2006ApJ...653..861M}. \cite{2007MNRAS.382..801M} found no environmental effect on the stellar mass--metallicity relation and conclude that the evolution of the galaxies is largely independent of their environments. They do find marginal increase in the chemical enrichment level at a fixed stellar mass in denser environments, so environments can play a moderate role in galaxy evolution. An alteration in the TFR in denser environments could in theory be due to a temporary increase/accelerated star formation or to photometric or kinematic asymmetry due to dynamical interactions of galaxies. But we do not find strong evidence for this.

\item We have established a procedure for ascertaining if any trend in the SDSS-based TFRs with a variable of interest is due to physics or sample bias. In general, it is difficult to convert the observed SDSS-based TFRs as a function of $v_{\rm FWHM}$ to TFRs as functions of $v_{\rm circ}$ in a quick, simple manner because there is a large amount of scatter between $v_{\rm FWHM}$ and $v_{\rm circ}$ due to aperture biases, even at higher redshifts where the apparent sizes of galaxies are smaller. Instead, it is most useful to work in magnitude versus $v_{\rm FWHM}$ space and convert an assumed TFR as a function of $v_{\rm circ}$ into this regime with the aid of simulations of $v_{\rm circ}$ versus $v_{\rm FWHM}$ for the given size distribution of the galaxy sample. Looking at the SDSS line width-based TFR in multiple bands proves to be very useful because given two groups of galaxies which may have disparities in their $v_{\rm circ}$ versus $v_{\rm FWHM}$ relationship due to aperture biases, these disparities (e.g. steepening) should be reflected in the same way in the SDSS-based TFRs in all bands assuming the TFR is the same for both groups.

\item Using a steeper TFR for late type galaxies, like that of \cite{2006ApJ...653..861M}, is also consistent with the data. The observed SDSS $v_{\rm FWHM}$-based slopes are still expected to steepen approximately $7$~per~cent from the lowest to the highest density groups due to aperture bias effects. Two different sloped TFRs (as function of $v_{\rm circ}$), if their intercepts are carefully chosen, can predict the same slope and intercept for the observed SDSS $v_{\rm FWHM}$-based TFR. But they would also predict a statistically different distribution of $v_{\rm FWHM}$, which we do not observe across density groups, so both the TFRs of \cite{2006ApJ...653..861M} and \cite{2007AJ....134..945P} cannot be consistent at the same time across different density groups. The choice of a particular TFR does not affect our ability probe for differences in the inferred TFRs across density.
\end{itemize}

\section*{Acknowledgments}
We would like to thank Chris Blake for discussions on statistical analysis we used in this paper. 
We would also like to thank the anonymous referee, who provided us with constructive comments and suggestions, particularly on addressing effects which may limit the sensitivity of the technique.
PM and MM acknowledge support from Swinburne University Centre for Astrophysics and Supercomputing (CAS) Vacation Scholarships. We also acknowledge support from the Australian Research Council Discovery Project DP1094370.

Funding for SDSS-III has been provided by the Alfred P. Sloan Foundation, the Participating Institutions, the National Science Foundation, and the U.S. Department of Energy Office of Science. The SDSS-III web site is http://www.sdss3.org/.

SDSS-III is managed by the Astrophysical Research Consortium for the Participating Institutions of the SDSS-III Collaboration including the University of Arizona, the Brazilian Participation Group, Brookhaven National Laboratory, University of Cambridge, University of Florida, the French Participation Group, the German Participation Group, the Instituto de Astrofisica de Canarias, the Michigan State/Notre Dame/JINA Participation Group, Johns Hopkins University, Lawrence Berkeley National Laboratory, Max Planck Institute for Astrophysics, New Mexico State University, New York University, Ohio State University, Pennsylvania State University, University of Portsmouth, Princeton University, the Spanish Participation Group, University of Tokyo, University of Utah, Vanderbilt University, University of Virginia, University of Washington, and Yale University.

\bibliography{mybib}{}

\appendix

\section{Model for SDSS-observed H$\alpha$ profile}\label{sec:model}

\begin{figure}
\centering
\includegraphics[width=0.47\textwidth]{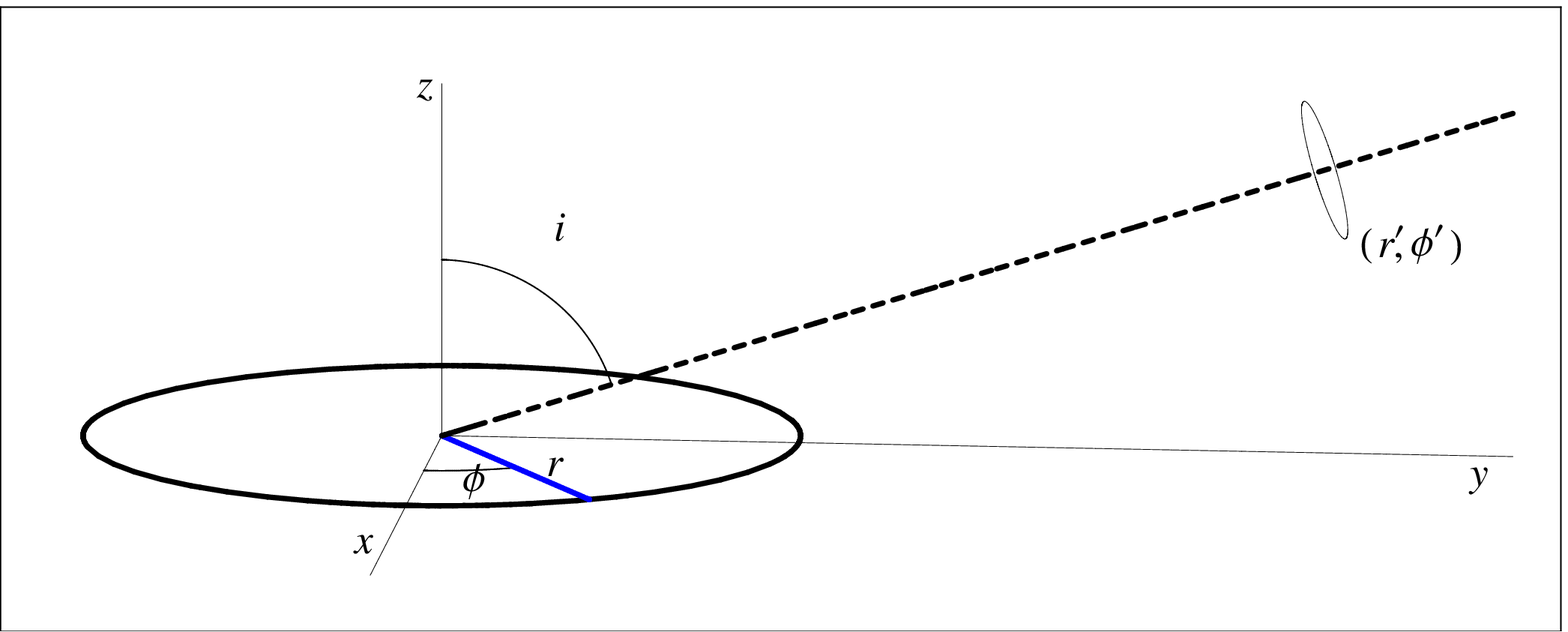}
\caption{Coordinate system used for our model of a disc galaxy. The coordinates $(r,\phi)$ describe a point in the plane of the disc and $i$ is the observed inclination angle. The circle on the right hand side, perpendicular to the line-of-sight, is meant to indicate the $3''$ aperture. Coordinates $(r^\prime,\phi^\prime)$ are in the plane of the fibre, with the centre of the fibre ($r^\prime=0$) pointing to the centre of the galaxy ($r=0$).}
\label{fig:schema}
\end{figure}

We describe here the simple model used to simulate synthetic populations and understand the correlation between the SDSS-measured line widths and true disc velocities. In our model, we assume an infinitely thin disc with an inclination angle $i$ to the observer's line-of-sight. The polar coordinates $(r,\phi)$ locate positions measured in the disc. A diagram of the coordinate system is presented in Figure~\ref{fig:schema}.

RCs of disc galaxies can often be well-modelled by an $\arctan$ function \citep{2007AJ....134..945P} 
(although we consider the effect of rising and falling profiles as well in Appendix~\ref{sec:RC}).
 The rotational velocity as a function from the centre of the disc is:
\begin{equation}
v(r)=\frac{2}{\pi}v_{\rm circ} \arctan \left(\frac{r}{r_{\rm t}}\right)
\end{equation}
where $v_{\rm circ}$ is the asymptotic circular velocity and $r_{\rm t}$ is the turn-over radius (the distance where $v(r)=v_{\rm circ}/2$). Then, the observed velocity at $(r,\phi)$ is:
\begin{equation}
v_{\rm obs}(r,\phi)=v(r)\sin(i)\cos(\phi)
\end{equation}
assuming we have subtracted off the systemic velocity of the galaxy.

We assume a simple exponential surface brightness profile for the disc given by:
\begin{equation}
I(r)=I_0\exp \left(-\frac{r}{h}\right)
\end{equation}
where $I_0$ is the central surface brightness and $h$ is the exponential scale radius. The intensity observed from an infinitesimal element $r\,dr\,d\phi$ is $I(r)r\,dr\,d\phi$.

The spectrum observed from an infinitesimal element $r\,dr\,d\phi$ will be centered at velocity $v_{\rm obs}(r,\phi)$. We assume a Gaussian profile for the emission with velocity dispersion (standard deviation) $\sigma=20$~km~s$^{-1}$ (based on the typical value as observed for spiral type galaxies in \cite{2008MNRAS.390..466E} and \cite{2010MNRAS.401.2113E}, which may range between $10$--$30$~km~s$^{-1}$). The dispersion $\sigma$ is the galaxy's velocity dispersion (for H$\alpha$) at an infinitesimal area in the disc, which we assume is constant as a function of $r$. The shape of the emission profile for element $r\,dr\,d\phi$ is then:
\begin{equation}
P^\prime(v)=I(r)r\,dr\,d\phi \times N(v;v_{\rm obs}(r,\phi),\sigma)
\end{equation}
where the notation $N(x;\mu,\sigma)$ means a normal distribution as a function of $x$ with mean $\mu$ and standard deviation $\sigma$.

To calculate the full H$\alpha$ emission profile of the galaxy, the following integration can be performed, numerically, over the entire disc:
\begin{equation}
P_{\rm full}(v)=\displaystyle\int_{\phi=0}^{2\pi} \int_{r=0}^\infty   I(r) N(v;v_{\rm obs}(r,\phi),\sigma) r\,dr\,d\phi. 
\end{equation}

To calculate the profile as observed by a $3''$ aperture, a similar integral can be evaluated numerically:
\begin{equation}
P_{\rm full}(v)=\displaystyle\int_{\phi=0}^{2\pi} \int_{r=0}^\infty   I(r) N(v;v_{\rm obs}(r,\phi),\sigma) F(r,\phi) r\,dr\,d\phi 
\end{equation}
where the added function $F(r,\phi)$ describes the fraction of the emission from $(r,\phi)$ that enters the $3''$ fibre. Establish new coordinates $(r^\prime,\phi^\prime)$ to represent points in the plane $P$ perpendicular to the line-of-sight with $r=0$ projecting onto $r^\prime =0$. Then the fibre can then be described as a circle $C$ of radius $1.5''$ in plane $P$. Supposing that the fibre is centered on the galaxy's centre, $r=0$, the fibre's centre is then at $r^\prime =0$. Without accounting for seeing, only points in the galaxy's disc that project to within circle $C$ will contribute to the observed emission profile. A point $(r,\phi)$ projects to plane $P$ to have $r^\prime={\rm proj}(r,\phi)$
where
\begin{equation}
{\rm proj}(r,\phi)=r\sqrt{\cos^2(\phi) + \sin^2(\phi)\cos^2(i)}.
\label{eqn:proj}
\end{equation}
Thus, without accounting for seeing, the function $F(r,\phi)$ is:
\begin{equation}
F_{\rm no~seeing}(r,\phi) = \begin{cases}
1 & {\rm proj}(r,\phi) \leq 1.5'', \\
0 & {\rm proj}(r,\phi) >    1.5''. 
\end{cases}
\end{equation}

\begin{figure}
\centering
\includegraphics[width=0.47\textwidth]{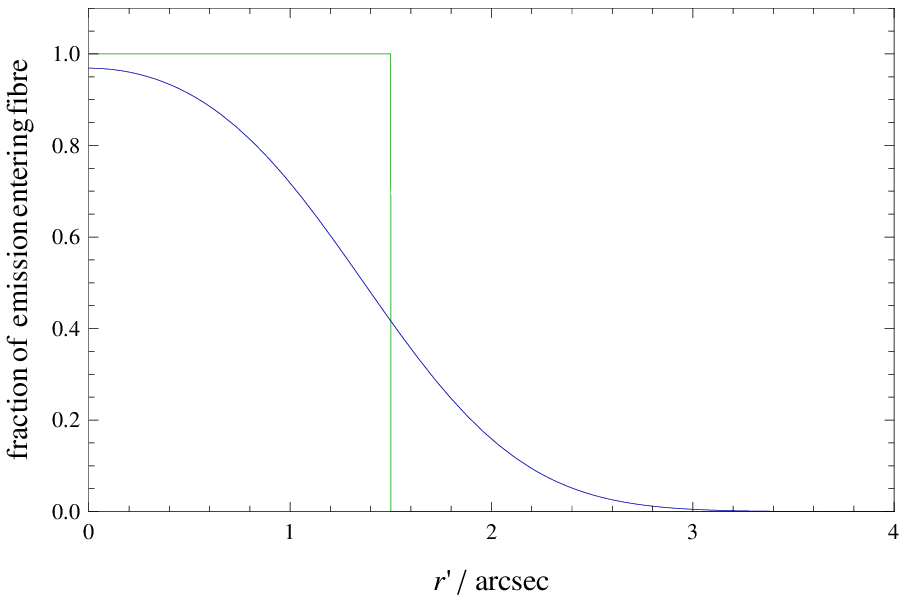}
\caption{Figure showing the fraction of projected light from the galaxy to a distance $r^\prime$ from the fibre's centre that enters the $3''$ diameter fibre, without accounting for seeing (green) and with $1.4''$ FWHM seeing (blue).}
\label{fig:seeing}
\end{figure}

To account for seeing, we assume a Gaussian convolution kernel with FWHM of $1.4''$. That is, the light that projects onto $(r^\prime,\phi^\prime)$ is redistributed as a 2D Gaussian. The fraction of this light that will enter the fibre is then the volume under the 2D Gaussian integrated over the surface with boundary $C$. This fraction only depends on $r^\prime$, the distance from the centre of the fibre, and not $\phi^\prime$. The integral is preformed numerically and the resulting fraction is shown in Figure~\ref{fig:seeing}. The fraction (blue line in Figure~\ref{fig:seeing}) can be approximated as:
\begin{equation}
f(r^\prime)=\frac{1}{2}\left( \rm{Erf}(1.5243 - 1.1157 r^\prime) + \rm{Erf}(1.5243 + 1.1157 r^\prime) \right)
\label{eqn:f}
\end{equation}
where
\begin{equation}
\rm{Erf}(z) = \frac{2}{\sqrt{\pi}} \displaystyle\int_0^z \exp(-t^2) \,dt.
\end{equation}

Hence, the function $F(r,\phi)$, which takes into account seeing, is the following:
\begin{equation}
F(r,\phi) = f({\rm proj}(r,\phi)),
\end{equation}
where $f$ is given by Equation~\ref{eqn:f} and ${\rm proj}(r,\phi)$ is given by Equation~\ref{eqn:proj}.

Adding $1.4''$ seeing does not have a strong effect on the measured FWHM of the emission profile. The difference between the FWHM with and without seeing is typically less than $2-5$~km~s$^{-1}$, due to the fibre ($3''$) being much larger than the seeing. This is useful to note because we have assumed a constant seeing while seeing will vary in the SDSS but it does not have a big effect on the measured line widths.

An example H$\alpha$ emission profile with and without the SDSS $3''$ aperture bias is presented in Figure~\ref{fig:TP} for parameter values $i=80^\circ$, $h=3''$, $r_{\rm t}=2$ and $v_{\rm circ}=200$~km~s$^{-1}$. The parameter $I_0$ is just a normalization constant, which does not effect the calculation of the FWHM of the observed profile.

\section{Effect of Systematics on Simulations}\label{sec:systematics}

\subsection{Effect of seeing on our results}\label{sec:seeing}

We have used a single seeing of $1.4''$ for all galaxies in our simulated of the observable fibre H$\alpha$ line width. However, the spectroscopic seeing for SDSS observations may range typically from $1''$ to $2''$. We run additional simulations of synthetic observations of $25,000$ galaxies with $1.1''$, $1.4''$, $1.7''$, $2.0''$ seeing. The relation between $v_{\rm circ}$ and $v_{\rm FWHM}$ also remains linear for galaxies with $z>0.045$ assuming any of these seeing, and we find that in the worst-case, the slope of the relation between $v_{\rm circ}$ and $v_{\rm FWHM}$ is changed by $1.3$~per~cent from $1.1''$ to $2.0''$. 

\subsection{Effect of rotation curve shape on our results}\label{sec:RC}

We have only considered the case where the RCs are asymptotically flat. However, disc galaxies have been observed to have rising, flat, or falling RCs (\citealt{2006ApJ...640..751C,1996MNRAS.281...27P}, see also erratum \citealt{1996MNRAS.283.1102P}). The RC profiles have been found to depend strongly on luminosity. Typically, the RCs may increase up to $0.8R_{\rm opt}$ (corresponding to $2.56h$ for a pure disc profile) and then decline for the most luminous galaxies or continue increasing for the less luminous ones. Since we cannot reconstruct RCs from SDSS data, we consider the effects different RCs could have on the observed fibre widths in our sample in extreme cases and show that they are negligible. We consider a galaxy with apparent turn-over radius $r_{\rm t}=0.5$'' and $h=1.5$'', which is in the small end of our sample. We suppose that the RC has an arctan profile out to $0.8R_{\rm opt}$ ($2.56h$), followed by either a flat profile or an upward/downward profile with slope $\Delta v/\Delta r = 10~{\rm km}~{\rm s}^{-1}$''. But the variation in the observed velocity width is found to be comparable to varying the seeing (the change in the width is $<10~{\rm km}~{\rm s}^{-1}$). And the variation in the observed width is virtually none for median-sized galaxies in our sample. This means that the SDSS fibre spectra are hardly sensitive to the shape of the RC beyond $0.8R_{\rm opt}$, where the profile may turn-over, remain flat, or rise. 

\subsection{Effect of Malmquist-type biases}\label{sec:malmquist}

Malmquist-type biases may be present in any TFR sample. These biases may refer to the fact that dimmer galaxies at higher redshift could be missing from the sample due to the flux limit of the sample, or flux errors scattering galaxies about the limit. These effects may weaken the observed slope of the TFR. The Tully-Fisher relations of \cite{2007AJ....134..945P} suffers from Malmquist bias, leading them to underestimate the true slope of the TFR. Since we have not corrected for this here, our $v_{\rm FWHM}$ TFR will also have a shallower slope, but, since our method is self consistent across the different environments, this weakening of the slope will not affect our conclusions.

\bsp
\label{lastpage}
\end{document}